\documentclass[referee]{raa}            
\usepackage{graphicx,times}             
\usepackage{gensymb}
\usepackage{float}

\begin{document}

   \title{Long-term $V(RI)_{c}$ CCD photometry of pre-main-sequence stars in the association Cepheus OB3$^*$
\footnotetext{$*$The photometric data will be available in electronic form via CDS VizieR Online Data Catalogue and by e-mail requested to S. Ibryamov at sibryamov@shu.bg.}
}

   \volnopage{Vol.0 (20xx) No.0, 000--000}      
   \setcounter{page}{1}          

   \author{Sunay Ibryamov
      \inst{1}
   \and Gabriela Zidarova
      \inst{1}
   \and Evgeni Semkov
      \inst{2}
   \and Stoyanka Peneva
      \inst{2}
   }

   \institute{Department of Physics and Astronomy, University of Shumen,
             115, Universitetska Str., 9700 Shumen, Bulgaria; {\it sibryamov@shu.bg}\\
        \and
             Institute of Astronomy and National Astronomical Observatory, Bulgarian Academy of Sciences, 72, Tsarigradsko Shose Blvd., 1784 Sofia, Bulgaria\\
   }

   \date{Received~~...; accepted~~...}

\abstract{
Results from optical CCD photometric observations of 13 pre-main-sequence stars collected during the period from February 2007 to November 2020 are presented.
These stars are located in the association Cepheus OB3, in the field of the young star V733 Cephei.
Photometric observations, especially concerning the long-term variability of the stars, are missing in the literature.
We present the first long-term $V(RI)_{c}$ monitoring for them, that cover 13 years.
Results from our study indicate that all of the investigated stars manifest strong photometric variability.
The presented paper is a part of our program for the photometric study of PMS stars located in active star-forming regions. 
}

   \authorrunning{S. Ibryamov, G. Zidarova, E. Semkov, S. Peneva}            
   \titlerunning{Long-term $V(RI)_{c}$ CCD photometry of PMS stars in the association Cepheus OB3}  

   \maketitle

%
%
\section{Introduction}           
\label{sect:intro}

Pre-main-sequence (PMS) stars are frequently found in associations with dense molecular clouds.
One such association is Cepheus OB3 (Cep OB3), which is located in the region between R.A. 22$^{h}$44$^{m}$ to 23$^{h}$08$^{m}$ and Dec. +61$\degree$ to +64$\degree$ (Blaauw et al.~\cite{blaa59}).
The distance to this association was determined as 725 pc by Blaauw et al.~(\cite{blaa59}) and as 900$\pm$100 pc by Moreno-Corral et al.~(\cite{more93}).
The star-formation process in Cep OB3 was detailed studied by many authors (e.g. see Sargent~\cite{sarg79}, Moreno-Corral et al.~\cite{more93}, Naylor \& Fabian~\cite{nayl99}, Mikami \& Ogura~\cite{mika01}, Pozzo et al.~\cite{pozz03}, Allen et al.~\cite{alle12}, Huang et al.~\cite{huan14}).

One of the most important features of PMS stars is their photometric variability.
Both main classes of PMS stars $-$ the low-mass (M $\leq$ 2M$_{\odot}$) T Tauri stars (TTSs) and the more massive (2M$_{\odot}$ $\leq$ M $\leq$ 8M$_{\odot}$) Herbig Ae/Be stars (HAEBESs) exhibit various types of light variations.
Generally, TTSs are separated into two subgroups: classical T Tauri stars (CTTSs) and weak-line T Tauri stars (WTTSs).
Most of the features that characterize each subgroup suggest that spacious circumstellar accretion disks surround CTTSs, whereas they must have almost disappeared (at least their inner parts) in WTTSs (M\'{e}nard \& Bertout~\cite{mena99}).

CTTSs often show irregular variability (with photometric amplitudes up to 2-3 mag), associated with variable accretion rate from the circumstellar disk onto the stellar surface.
The brightness variations of CTTSs also might be due to rotating spots on the stellar surface (Herbst et al.~\cite{herb07}).
A classification scheme for CTTSs based on the light curve shape was proposed by Ismailov~(\cite{isma05}).
The photometric variability of WTTSs is due to cool spots or groups of spots on the stellar surface.
Amplitudes of this variability are about 0.03-0.3 mag, but in extreme cases can reach 0.8 mag in the $V$-band.

In some PMS stars, large amplitude dips in the brightness (reaching up to 3 mag in the $V$ band) are observed.
These dips last from days to some weeks or months, which presumably result from circumstellar dust or clouds obscuration (see Voshchinnikov~\cite{vosh89}, Grinin et al.~\cite{grin91}, Herbst et al.~\cite{herb07}).
The prototype of this group of PMS objects named UXors is UX Orionis (Grinin et al.~\cite{grin91}).

The stars included in the present study were selected from the International Variable Star Index (VSX) database of AAVSO (https://www.aavso.org/vsx/; visited in June 2020) by exact object type (YSO) and with the first condition that their location is within 25 arcmin of the typical FU Ori star V733 Cep (see Reipurth et al.~\cite{reip07}, Semkov \& Peneva~\cite{semk08}, Peneva et al.~\cite{pene10}), because our observations were carried out in the field centered on this star.
The second condition was the stars to be bright enough for the telescopes and cameras used.

Photometric observations, especially concerning the long-term variability of the monitored stars, are missing in the literature.
We present the first long-term $V(RI)_{c}$ photometry for these stars that cover 13 years.
The long-term observations are important for the exact classification and clarifying the nature of the young stars.
Some recent studies of PMS stars based on such observations were made by Rigon et al.~(\cite{rigo17}), Hamb\'{a}lek et al.~(\cite{hamb19}), Siwak et al.~(\cite{siwa19}), Semkov et al.~(\cite{semk19}), Evitts et al.~(\cite{evit20}), Froebrich et al.~(\cite{froe20}), Ibryamov et al.~(\cite{ibry20}), etc.

Section 2 in the present paper gives information about the telescopes and cameras used.
Section 3 describes the obtained results and their analysis.
Section 4 provides the conclusion.

\section{Observations and data reduction}
\label{sect:Obs}

We obtained photometric observations of the field centered on V733 Cep within the association Cep OB3.
The $V(RI)_{c}$ data presented in this paper were collected in the period from February 2007 to November 2020 with the 50/70-cm Schmidt and the 60-cm Cassegrain telescopes administered by the Rozhen National Astronomical Observatory (Bulgaria).
Only one of the stars (2MASS J22533430+6236307) is also visible in the field of view of the 1.3-m Ritchey-Chr\'{e}tien (RC) telescope of the Skinakas Observatory\footnote{Skinakas Observatory is a collaborative project of the University of Crete, the Foundation for Research and Technology, Greece, and the Max-Planck-Institut f{\"u}r Extraterrestrische Physik, Germany.} of the University of Crete (Greece), and we have $B$ data for this star.

The observations were performed with five different types of CCD cameras $-$ SBIG ST-8, SBIG STL-11000M and FLI PL16803 on the 50/70-cm Schmidt telescope, FLI PL09000 on the 60-cm Cassegrain telescope, and ANDOR DZ436-BV on the 1.3-m RC telescope.
Their technical parameters and specifications are given in Ibryamov et al.~(\cite{ibry15}).
All frames were taken through a standard Johnson$-$Cousins set of filters.
The photometric data were reduced by \textsc{idl} software package (standard \textsc{daophot} subroutine).
As a reference, the comparison sequence in the field around V733 Cep reported in Peneva et al.~(\cite{pene10}) was used.
All data were analyzed using the same aperture, which was chosen to have a 3 arcsec radius, while the background annulus was taken from 10 to 13 arcsec.

\section{Results and discussion}
\label{sect:Res}

The stars from our study are listed in Table~\ref{Tab:designations} in order of increasing right ascension.
The table contains the 2MASS designations, equatorial coordinates, classes, and $Gaia$ distances of the stars.
It can be seen from Table~\ref{Tab:designations} that 10 of the monitored stars are most probably members of the association Cep OB3.
There are no $Gaia$ distances to two of the stars (2MASS J22515547+6218170 and 2MASS J22543612+6243047), and 2MASS J22545815+6241327 likely is field star projected beyond the association because the distance to it seriously exceeds one determined to Cep OB3.
Roughly all stars (11) are included in the list of young stellar objects (YSOs): class II (PMS star with disk) published by Allen et al.~(\cite{alle12}).

{\footnotesize
\begin{table*}[h!]
  \caption{Designations, equatorial coordinates and classes of the stars from our study, and the $Gaia$ distances to them.}\label{Tab:designations}
  \begin{center}
  \begin{tabular}{cccccr}
	  \hline\hline
	  \noalign{\smallskip}
Nr. & 2MASS ID$^1$      & RA$_{J2000.0}$$^1$ & Dec$_{J2000.0}$$^1$ & Class$^2$ & $d$ [pc]$^3$ \\
1   & J22515547+6218170 & 22:51:55.47        & +62:18:17.0         & -         & -            \\
2   & J22533129+6237114 & 22:53:31.30        & +62:37:11.4         & II        & 807          \\
3   & J22533430+6236307 & 22:53:34.30        & +62:36:30.7         & II        & 819          \\
4   & J22534450+6238305 & 22:53:44.51        & +62:38:30.5         & II        & 884          \\
5   & J22541742+6236221 & 22:54:17.43        & +62:36:22.1         & II        & 898          \\
6   & J22543612+6243047 & 22:54:36.13        & +62:43:04.7         & II        & -            \\
7   & J22545815+6241327 & 22:54:58.15        & +62:41:32.7         & II        & 2624         \\
8   & J22550051+6233485 & 22:55:00.52        & +62:33:48.6         & II        & 843          \\
9   & J22550590+6234504 & 22:55:05.90        & +62:34:50.5         & II        & 947          \\
10  & J22551206+6228169 & 22:55:12.06        & +62:28:16.9         & II        & 897          \\
11  & J22551996+6218162 & 22:55:19.96        & +62:18:16.3         & -         & 867          \\
12  & J22552178+6237535 & 22:55:21.79        & +62:37:53.6         & II        & 833          \\
13  & J22552445+6239137 & 22:55:24.46        & +62:39:13.8         & II        & 953          \\
   	  \hline \hline
  \end{tabular}
  \end{center}
  {\textbf{References:} $^1$2Micron All-Sky Survey (2MASS) (Skrutskie et al.~\cite{skru06}); $^2$Allen et al.~(\cite{alle12}); $^3$$Gaia$ mean distance (Bailer-Joines et al.~\cite{bail18})}.\\
	\end{table*}} 

\begin{figure*}[h!]
	\centering
	\includegraphics[width=10cm, angle=0]{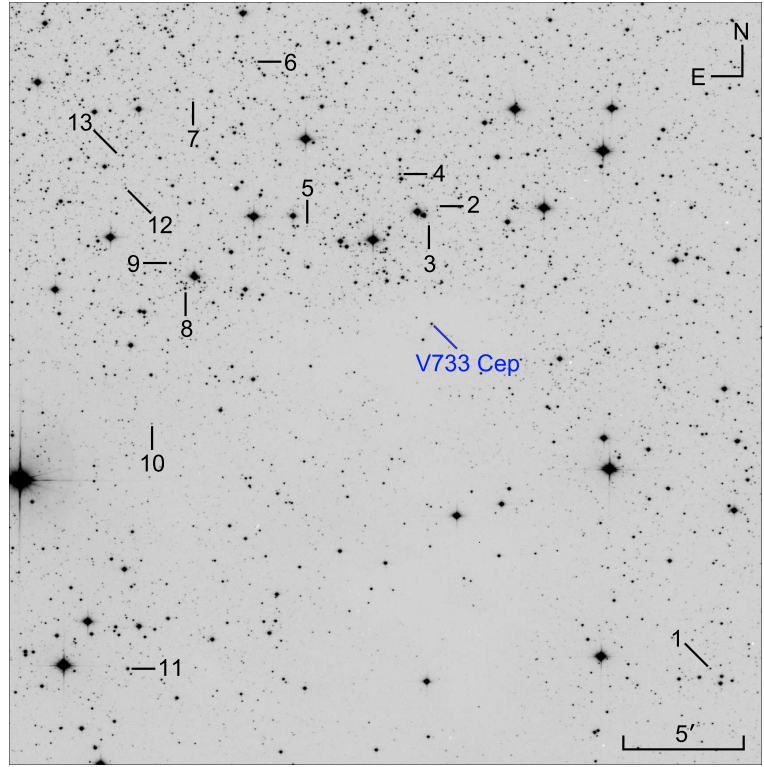}
	\caption{An $I_{c}$ band image of the vicinity of V33 Cep obtained with the 50/70-cm Schmidt telescope. The positions of the stars from our study and V733 Cep are marked. The stars are designated using their numbers given in Table~\ref{Tab:designations}.}\label{Fig:field}
\end{figure*}

Figure~\ref{Fig:field} features an $I_{c}$ band image of the vicinity of V733 Cep, where the positions of the stars from our study are marked.
The image was obtained with the 50/70-cm Schmidt telescope.

\begin{table}[h!]
	{\footnotesize
		\caption{The registered minimal and maximal magnitudes, and the amplitudes of variability in the $V(RI)_{c}$ bands of the stars from our study.}\label{Tab:amplitudes}
		\begin{center}
			\begin{tabular}{lcccccccccc}
				\hline\hline
				\noalign{\smallskip}
				Nr & Star (2MASS ID)   & $V_{min}$ & $V_{max}$ & $R_{c_{min}}$ & $R_{c_{max}}$ & $I_{c_{min}}$ & $I_{c_{max}}$ & $\Delta{V}$ & $\Delta{R_{c}}$ & $\Delta{I_{c}}$ \\ 		
				\noalign{\smallskip}
				\hline
				\noalign{\smallskip}
				1  & J22515547+6218170 & 18.82 & 17.72 & 17.39 & 16.32 & 15.83 & 14.95 & 1.10 & 1.07 & 0.88 \\
				2  & J22533129+6237114 & 19.66 & 18.51 & 18.69 & 17.11 & 16.87 & 15.64 & 1.15 & 1.58 & 1.23 \\
				3  & J22533430+6236307 & 20.17 & 17.93 & 18.62 & 16.51 & 16.82 & 15.12 & 2.24 & 2.11 & 1.70 \\
				4  & J22534450+6238305 & 17.38 & 15.09 & 16.01 & 13.76 & 14.71 & 12.85 & 2.29 & 2.25 & 1.86 \\
				5  & J22541742+6236221 & 19.38 & 18.82 & 19.48 & 16.89 & 18.26 & 15.66 & 0.56 & 2.59 & 2.60 \\
				6  & J22543612+6243047 & 19.01 & 18.14 & 18.89 & 16.77 & 17.01 & 15.28 & 0.87 & 2.12 & 1.73 \\
				7  & J22545815+6241327 & 19.18 & 18.22 & 18.22 & 17.08 & 16.49 & 15.67 & 0.96 & 1.14 & 0.82 \\
				8  & J22550051+6233485 & 17.12 & 15.45 & 16.14 & 14.15 & 14.74 & 13.06 & 1.67 & 1.99 & 1.68 \\
				9  & J22550590+6234504 & 16.76 & 15.92 & 15.94 & 14.62 & 14.50 & 13.25 & 0.84 & 1.32 & 1.25 \\
				10 & J22551206+6228169 & 19.46 & 17.98 & 18.11 & 16.92 & 16.45 & 15.70 & 1.48 & 1.19 & 0.75 \\
				11 & J22551996+6218162 & 14.93 & 14.20 & 14.39 & 13.47 & 13.51 & 12.75 & 0.73 & 0.92 & 0.76 \\
				12 & J22552178+6237535 & 18.49 & 17.03 & 17.45 & 15.85 & 16.19 & 14.68 & 1.46 & 1.60 & 1.51 \\
				13 & J22552445+6239137 & 19.29 & 18.39 & 19.36 & 17.18 & 17.54 & 15.77 & 0.90 & 2.18 & 1.77 \\
				\hline \hline
			\end{tabular}
	\end{center}}
\end{table}

The registered minimal and maximal magnitudes, and the amplitudes of variability in the $V(RI)_{c}$ bands of the stars are given in Table~\ref{Tab:amplitudes}.
It is important to mention that our observations were obtained mainly in the $R_{c}$ and $I_{c}$ bands.
That is why we have fewer $V$ observational points and the registered amplitudes for some stars in the $V$ band are smaller than these in $R_{c}$ and $I_{c}$ bands.
It can be seen from Table~\ref{Tab:amplitudes} that all stars exhibit brightness variations with large amplitudes.
The largest amplitude star is 2MASS J22541742+6236221 ($\Delta{I_{c}}$=2.60 mag).

\begin{figure}[h!]
	\includegraphics[width=8.3cm]{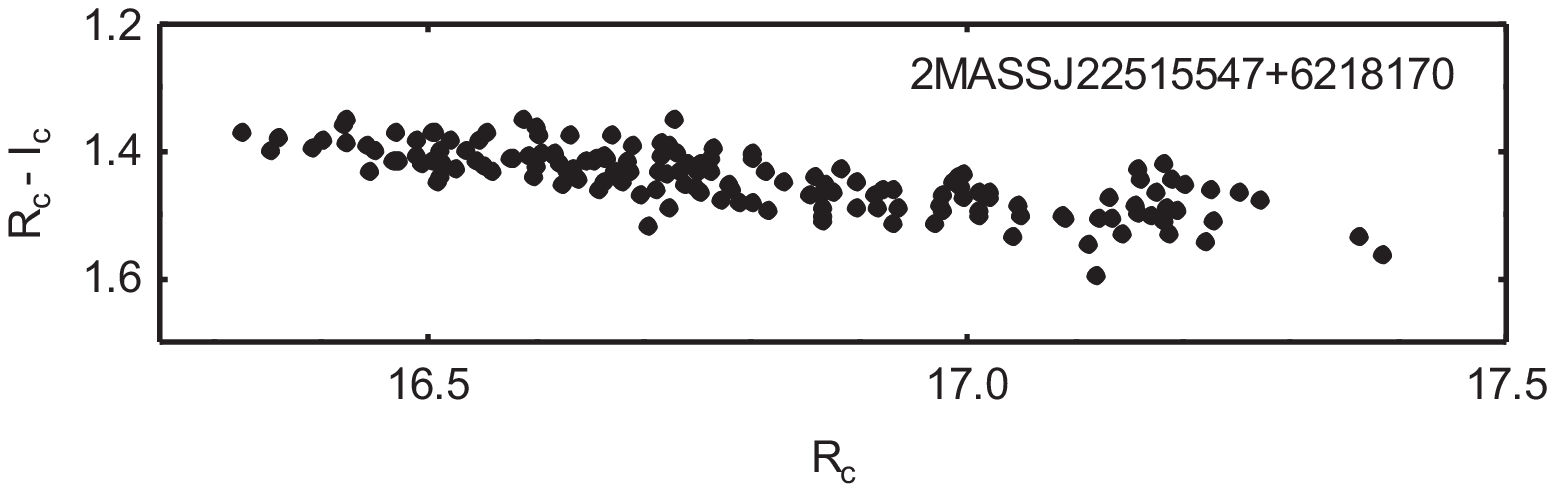}
	\includegraphics[width=8.3cm]{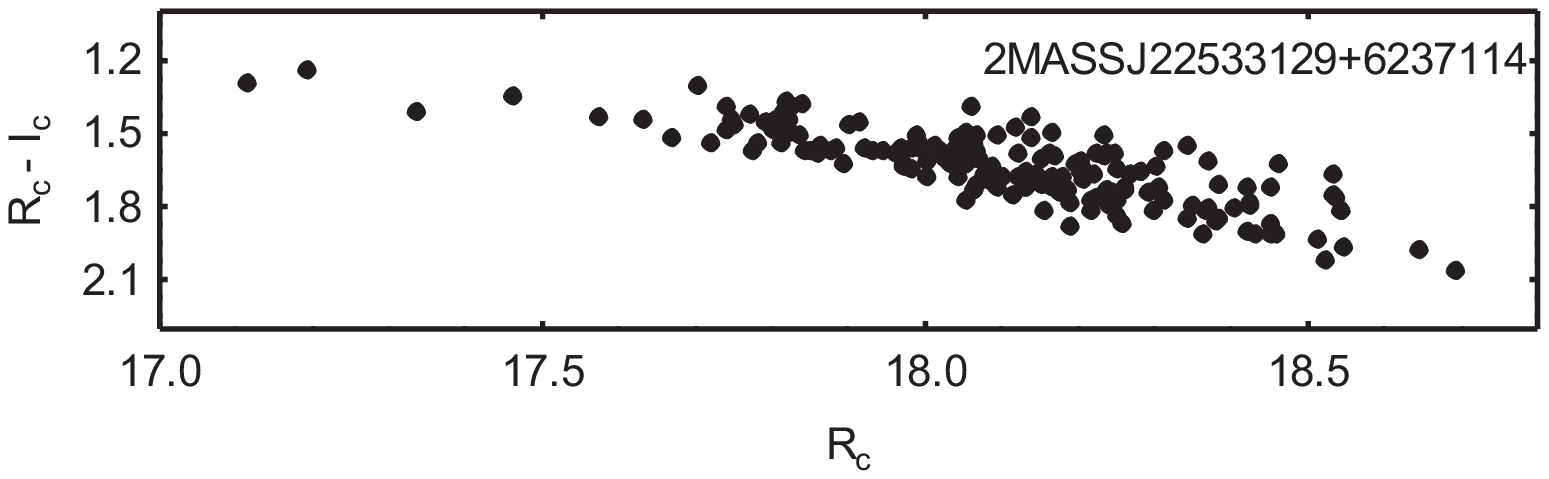}
	\includegraphics[width=8.3cm]{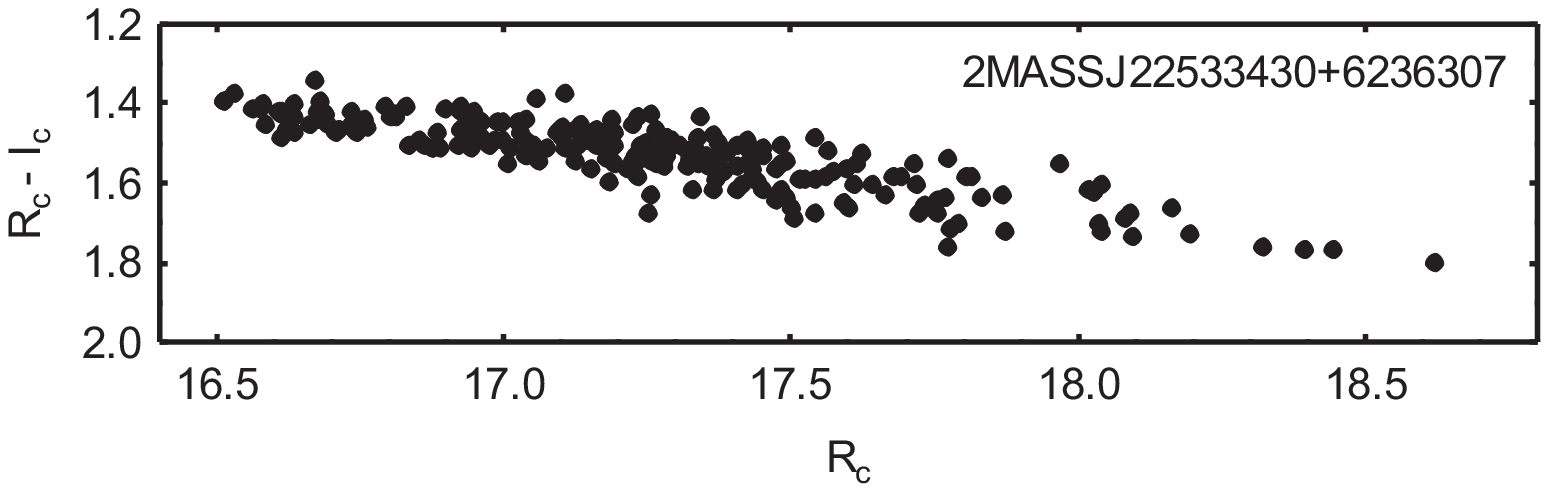}
	\includegraphics[width=8.3cm]{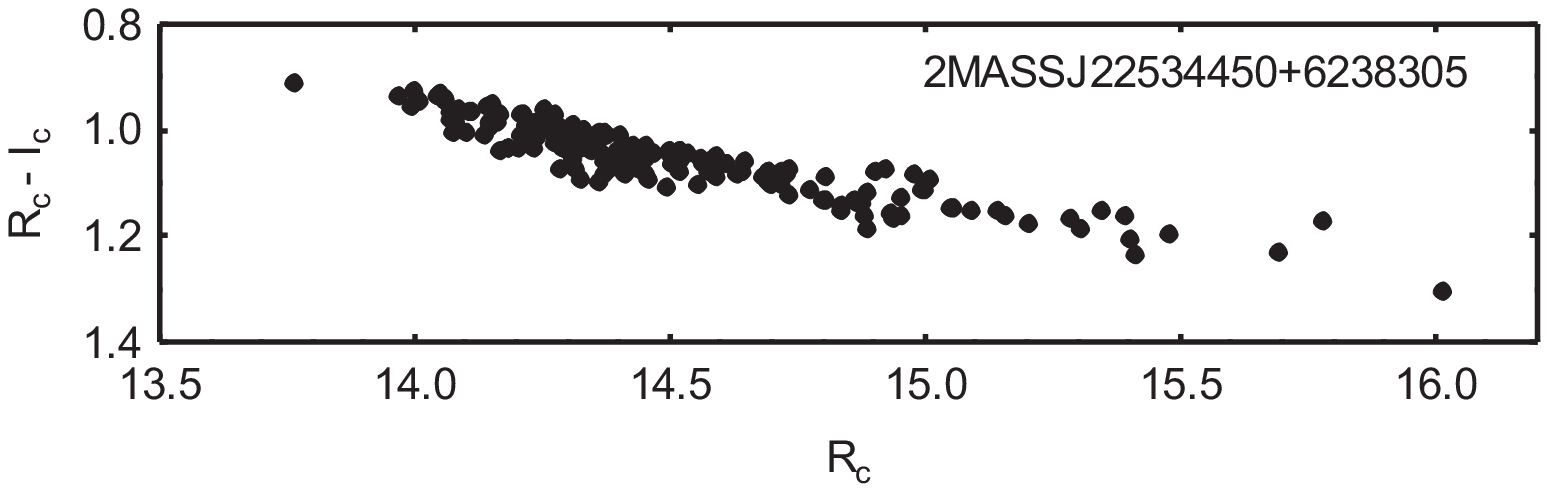}
	\includegraphics[width=8.3cm]{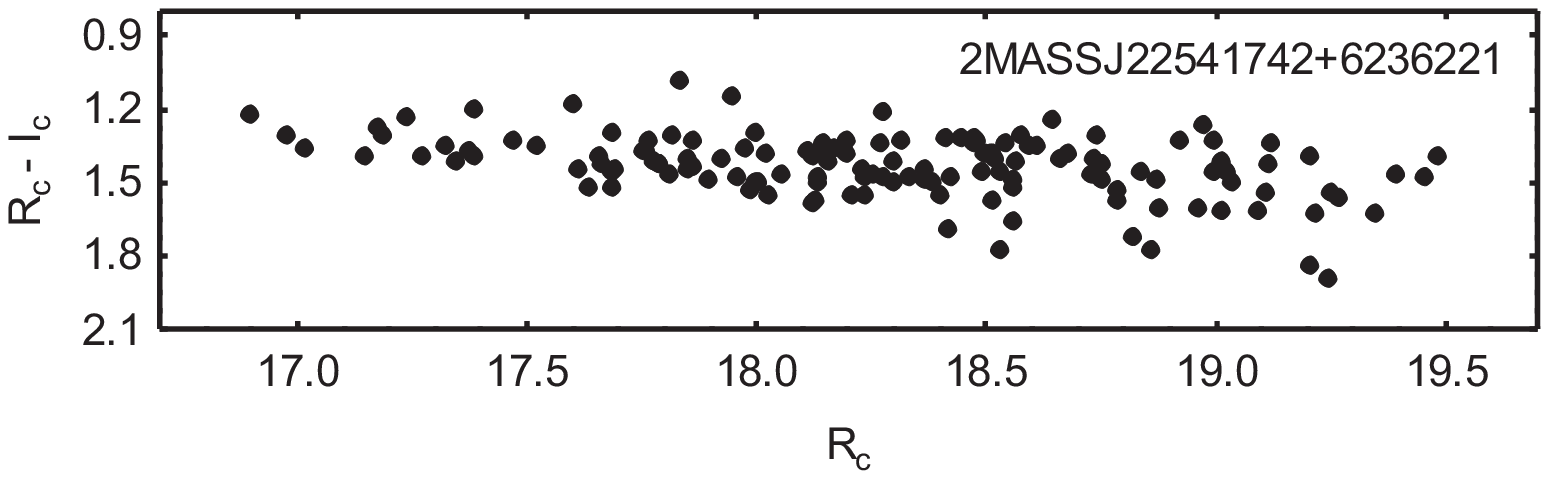}
	\includegraphics[width=8.3cm]{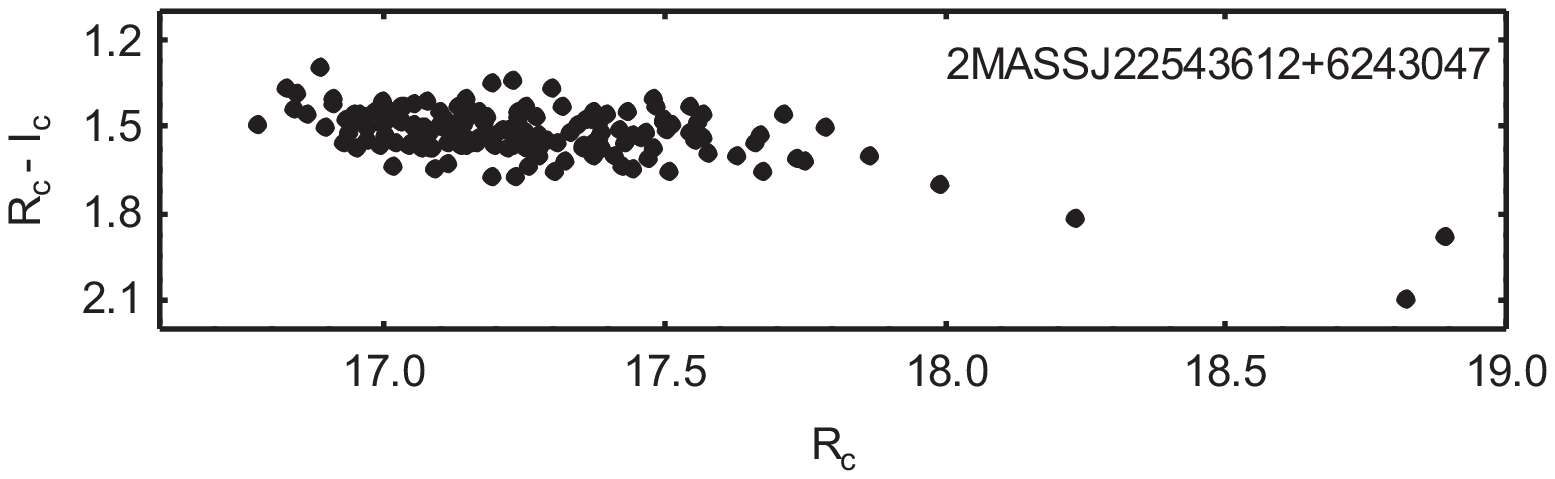}
	\includegraphics[width=8.3cm]{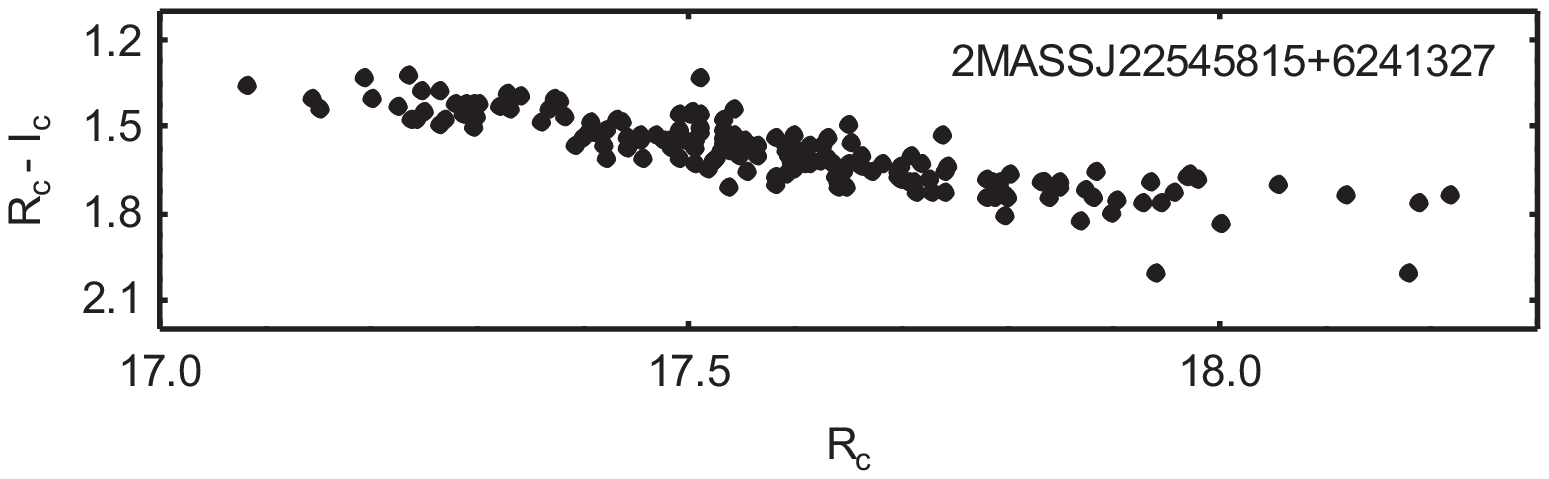}
	\includegraphics[width=8.3cm]{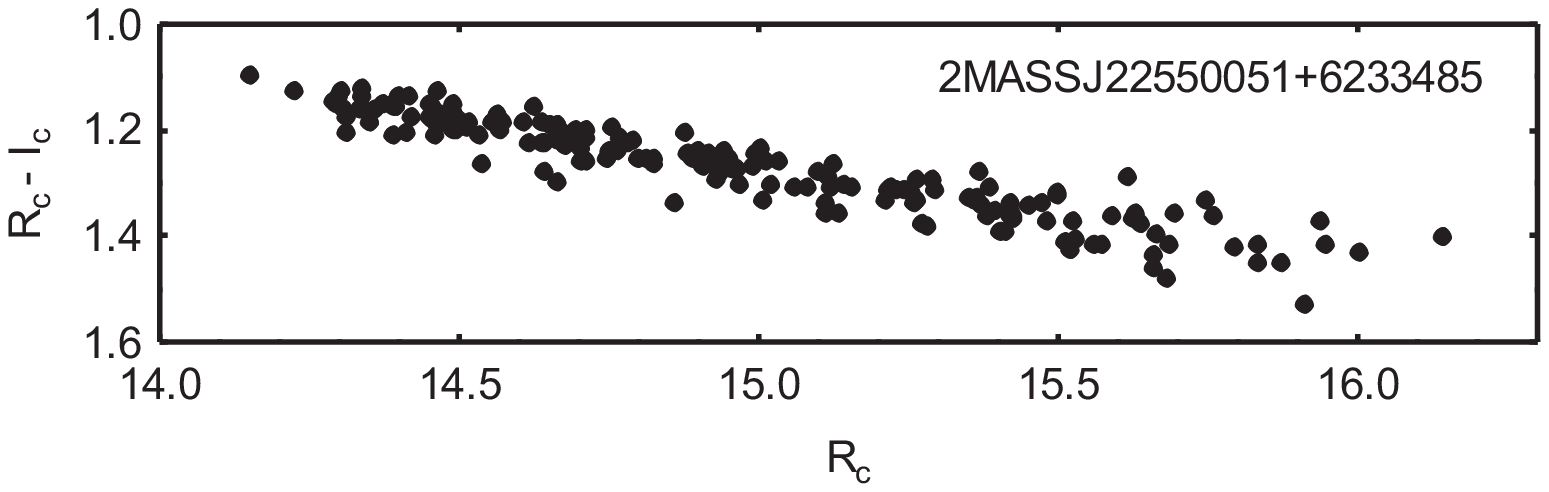}
	\includegraphics[width=8.3cm]{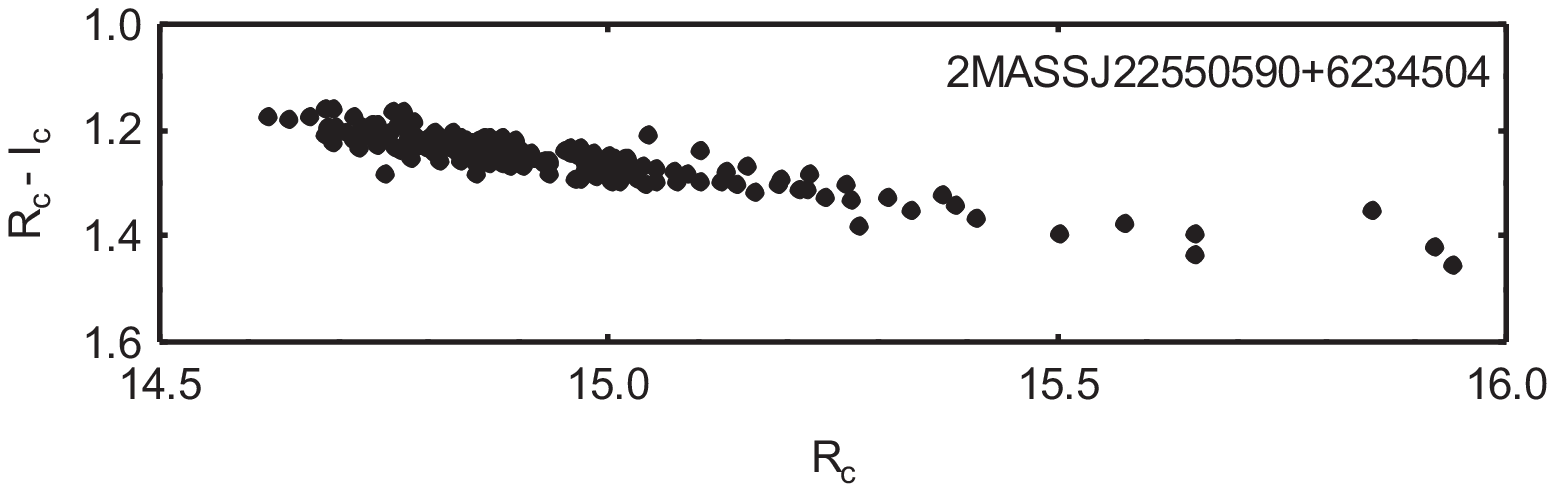}
	\includegraphics[width=8.3cm]{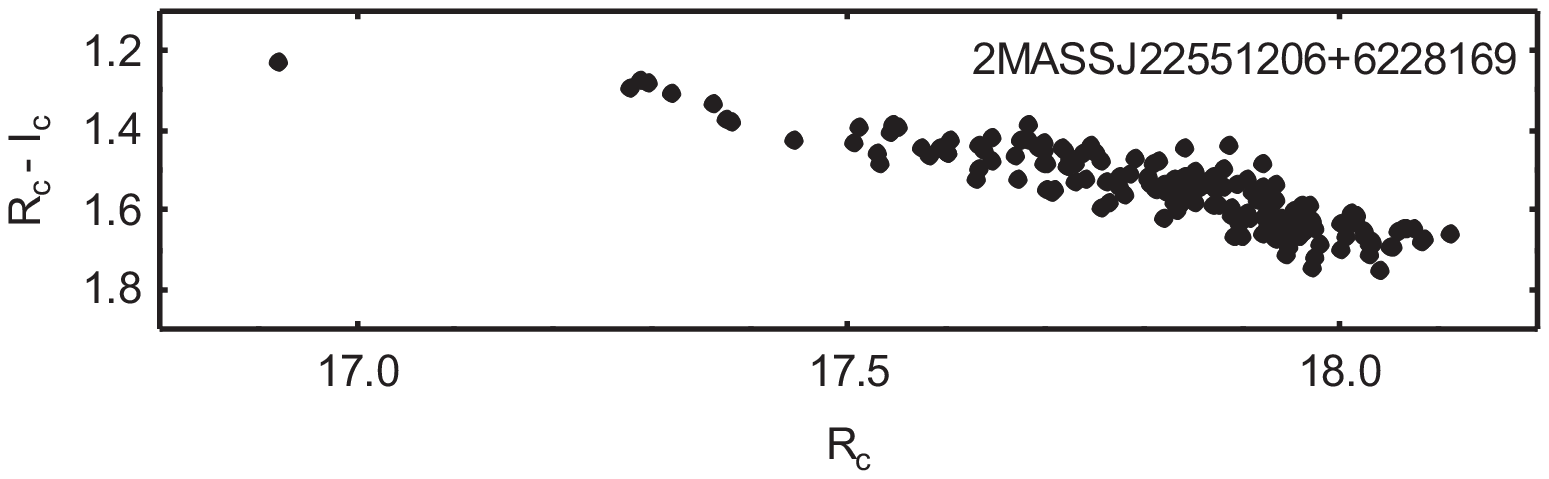}
	\includegraphics[width=8.3cm]{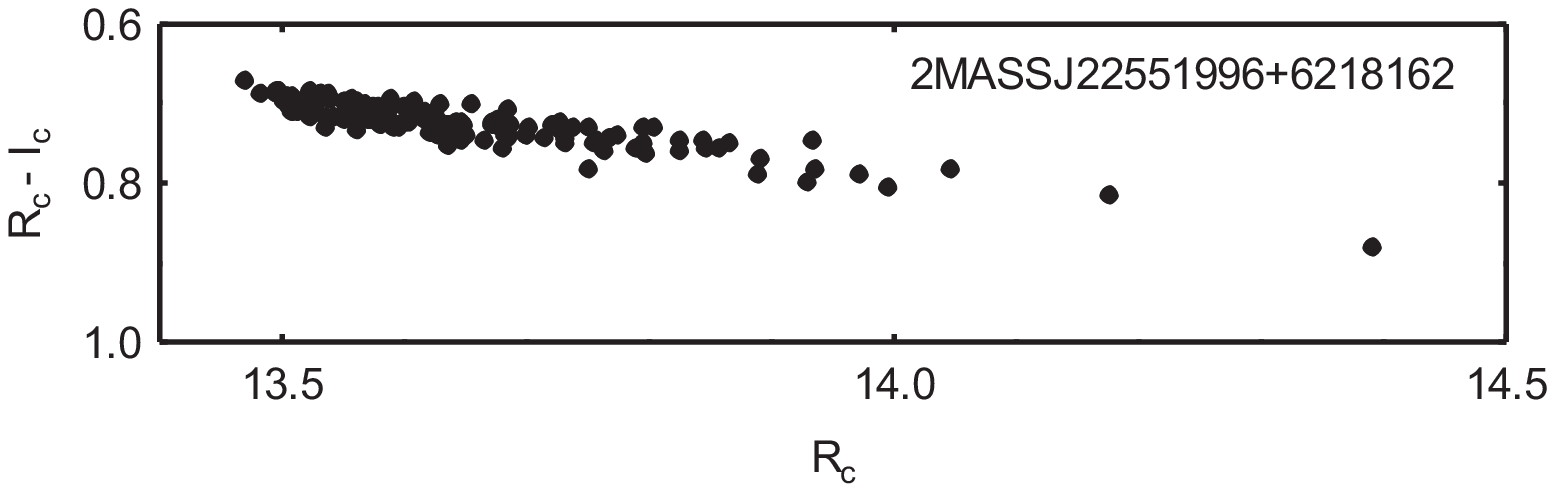}
	\includegraphics[width=8.3cm]{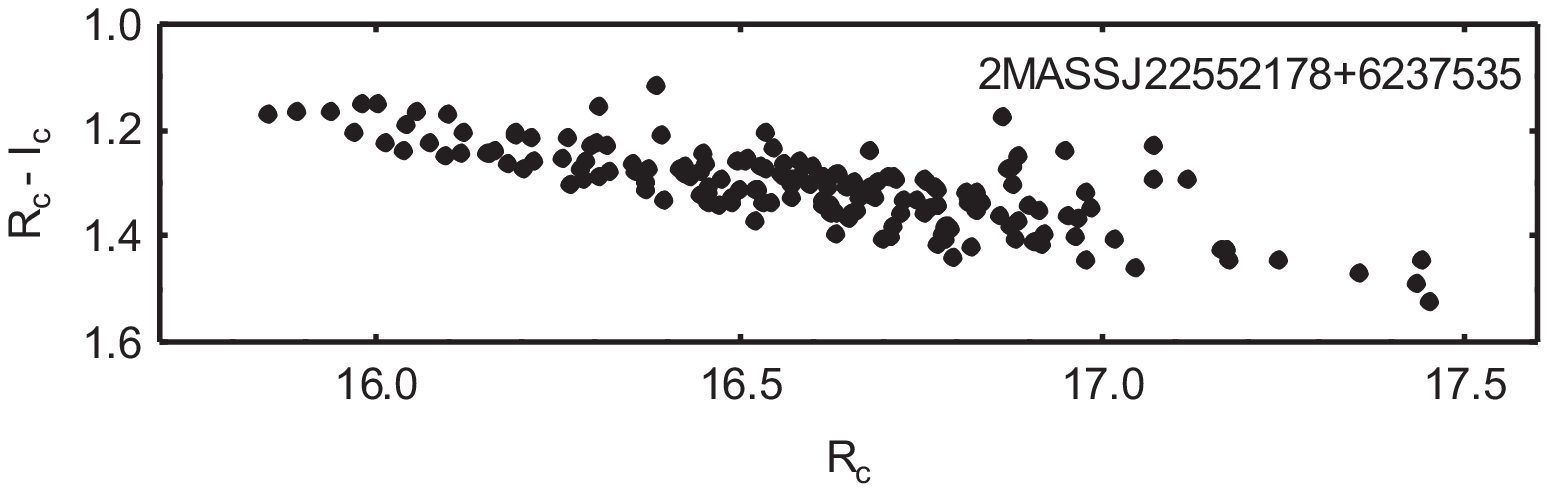}
	\includegraphics[width=8.3cm]{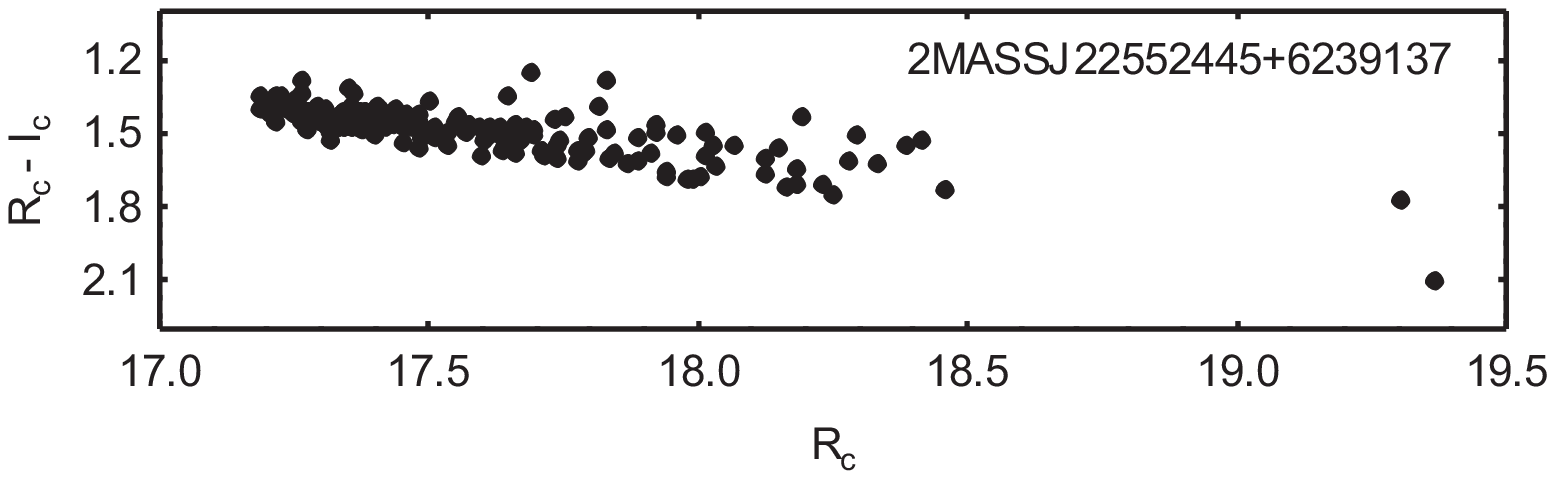}
	\caption{Color index $R_{c}-I_{c}$ versus the stellar $R_{c}$ magnitude of the stars from our study.}\label{Fig:colors}
\end{figure}

For all stars we constructed color-magnitude ($R_{c} - I_{c}$/$R_{c}$) diagrams, which are depicted in Fig.~\ref{Fig:colors}.
We used the software package \textsc{period04} (Lenz \& Breger~\cite{lenz05}) to search for periodicity in the light variations of the monitored stars.
In the following subsections, we discuss the stars in groups, that exhibit close photometric behavior.

\subsection{2MASS J22515547+6218170 and 2MASS J22545815+6241327}

\begin{figure}[h]
		\includegraphics[width=8.5cm]{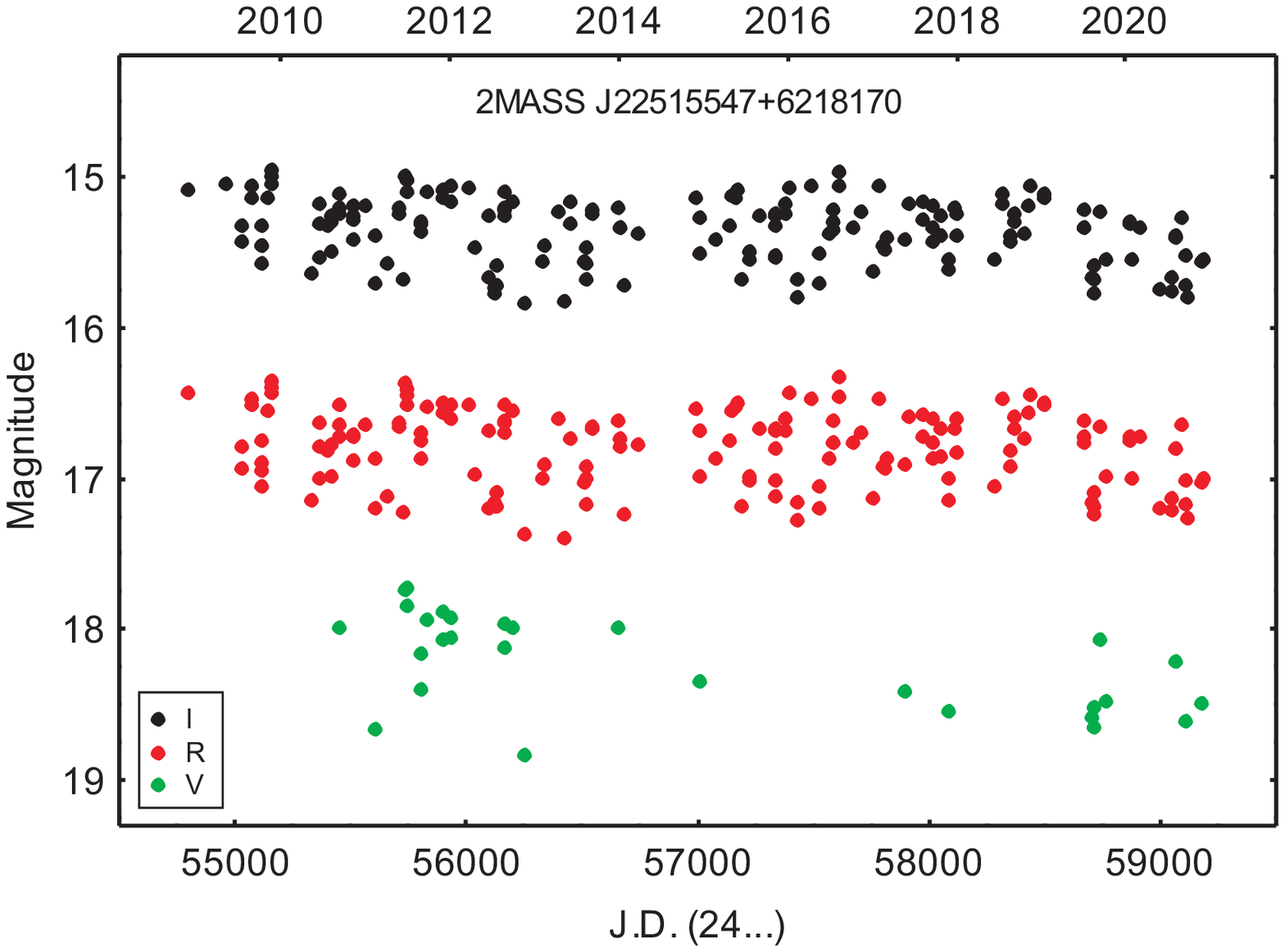}
		\includegraphics[width=8.5cm]{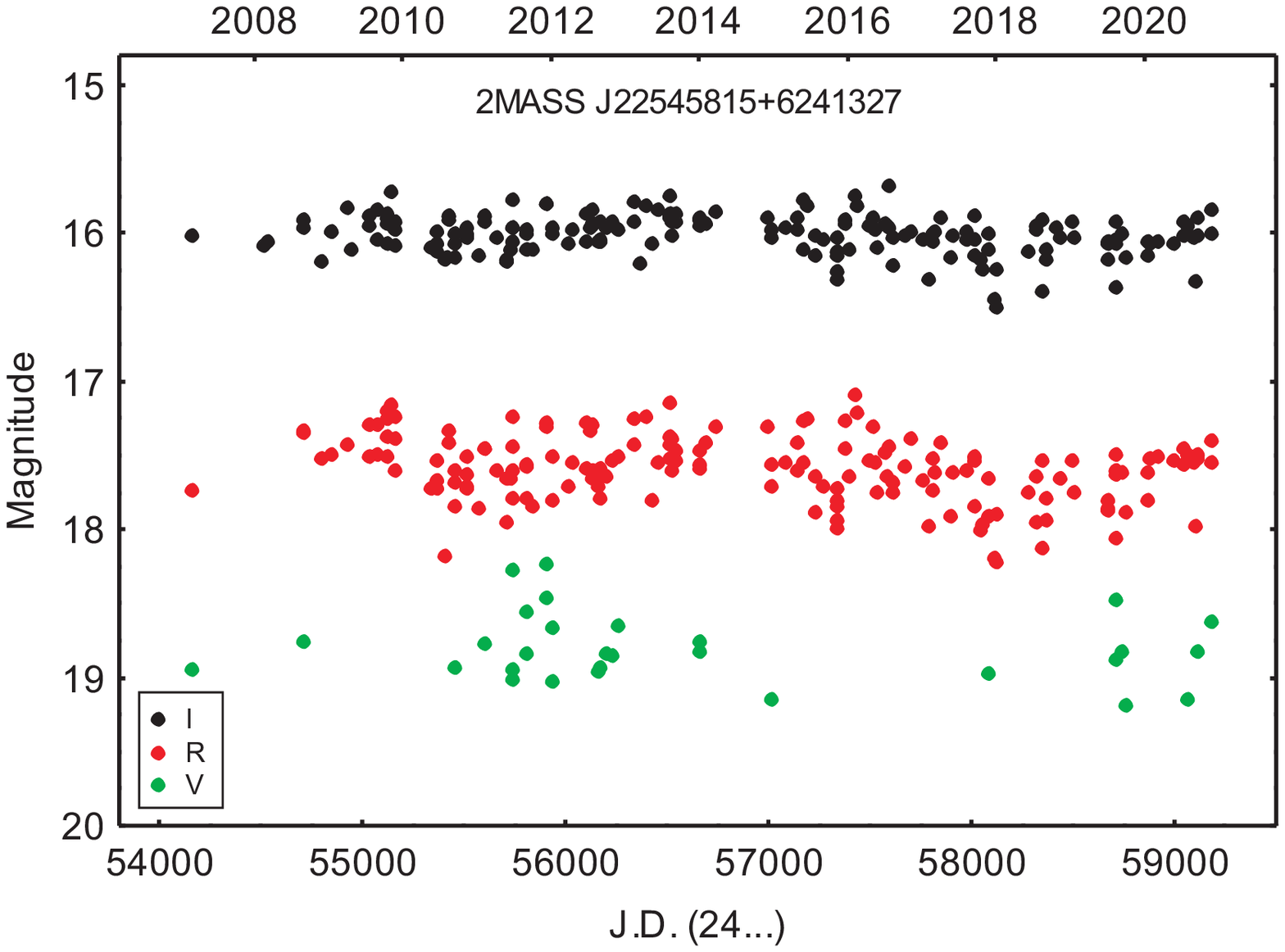}
		\caption{$V(RI)_{c}$ light curves of 2MASS J22515547+6218170 and 2MASS J22545815+6241327.}\label{Fig:curves1}
\end{figure}

During our monitoring, these stars exhibit rapid brightness variation, whose amplitude is almost the same in different years.
Their light curves constructed on the basis of our observations are shown in Fig.~\ref{Fig:curves1}.
The registered photometric amplitudes of the stars are given in Table~\ref{Tab:amplitudes}.
Usually, variability with such amplitudes is observed in CTTSs and it can be explained by rotational modulation of a spot or group of spots on the stellar surface, as well as by variable accretion rate.
Evidence of periodicity in the variability of 2MASS J22515547+6218170 and 2MASS J22545815+6241327 is not detected.

\subsection{2MASS J22533129+6237114 and 2MASS J22551206+6228169}

\begin{figure}[]
		\includegraphics[width=8.5cm]{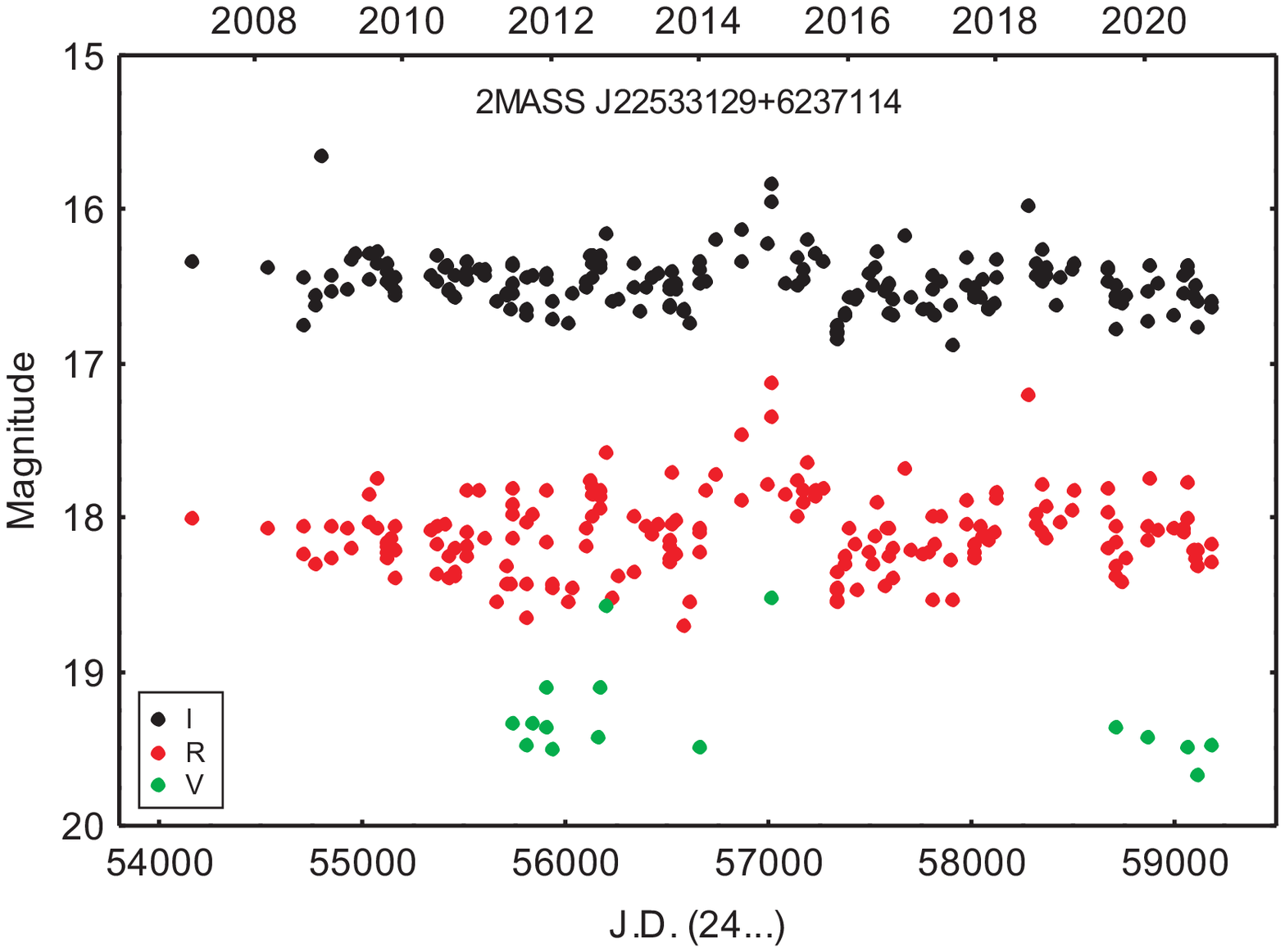}
		\includegraphics[width=8.5cm]{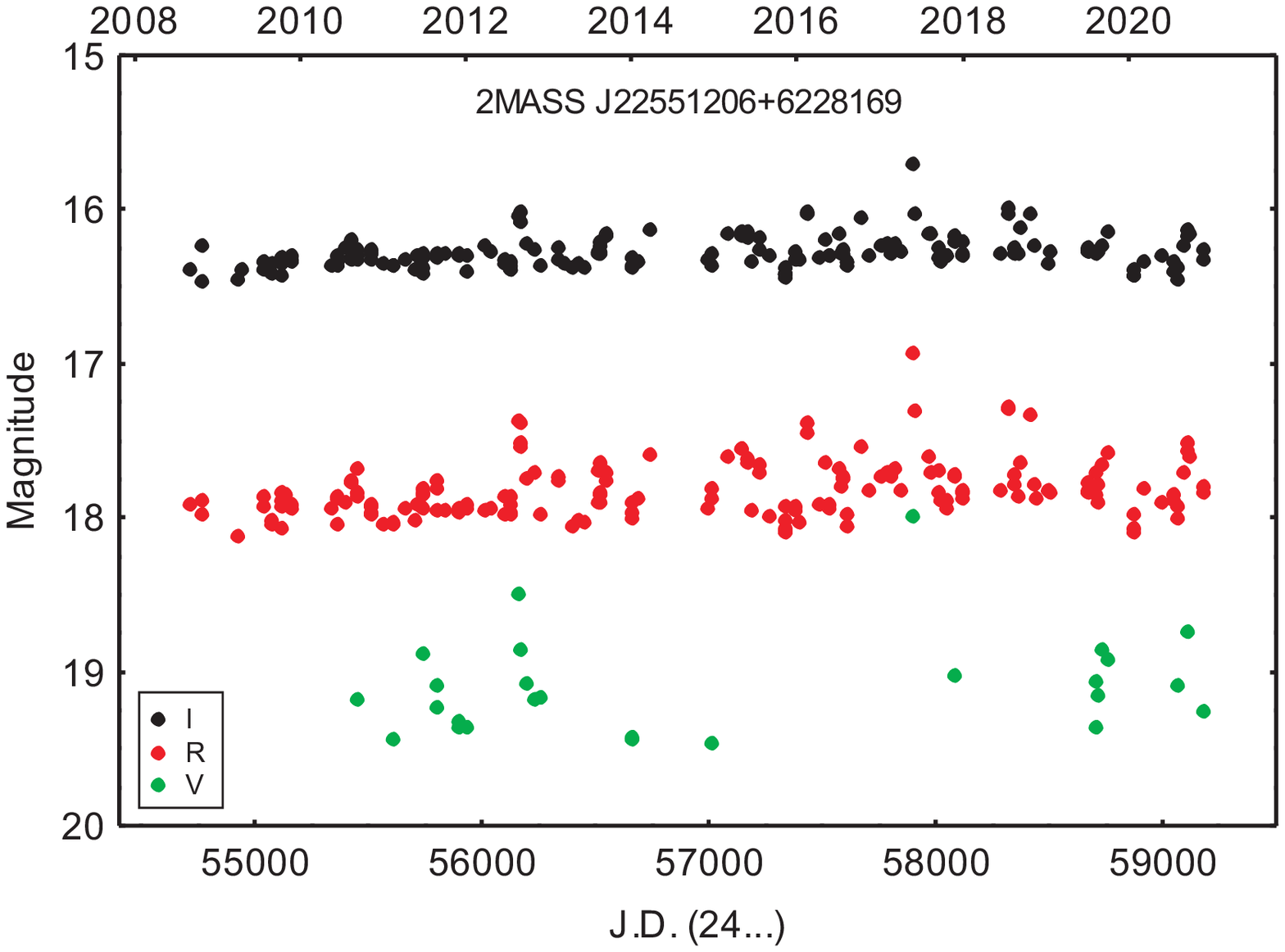}
		\caption{$V(RI)_{c}$ light curves of 2MASS J22533129+6237114 and 2MASS J22551206+6228169.}\label{Fig:curves2}
\end{figure}

These stars manifest photometric variability with sudden increases in brightness.
The light curves of the stars are displayed in Fig.~\ref{Fig:curves2}.
During our monitoring, clearly distinguishable flare events in the brightness of 2MASS J22533129+6237114 were registered in November 2008, December 2014 and June 2018.
Such events in the light variation of 2MASS J22551206+6228169 were observed in August 2012 and May 2017.
These non-periodic flares likely are caused by short-lived accretion-related events on the stellar surface.
The time periods of increases in the brightness are relatively short; usually, we have only one photometric point during the flares.

It can be seen from the color-magnitude diagrams (Fig.~\ref{Fig:colors}) that the stars become bluer when they increase the brightness.
Such color variation is also an indication for flare events.
Out of the flares, the variability of these stars is probably caused by rotating hot spots on the stellar surface and/or by variable accretion rate from the circumstellar disk.
Evidence of periodicity in the variability of 2MASS J22533129+6237114 and 2MASS J22551206+6228169 is not detected.

\subsection{2MASS J22533430+6236307, 2MASS J22534450+6238305, 2MASS J22543612+6243047, 2MASS J22550051+6233485, 2MASS J22550590+6234504, 2MASS J22551996+6218162 and 2MASS J22552445+6239137}

\begin{figure}[]
		\includegraphics[width=7.55cm]{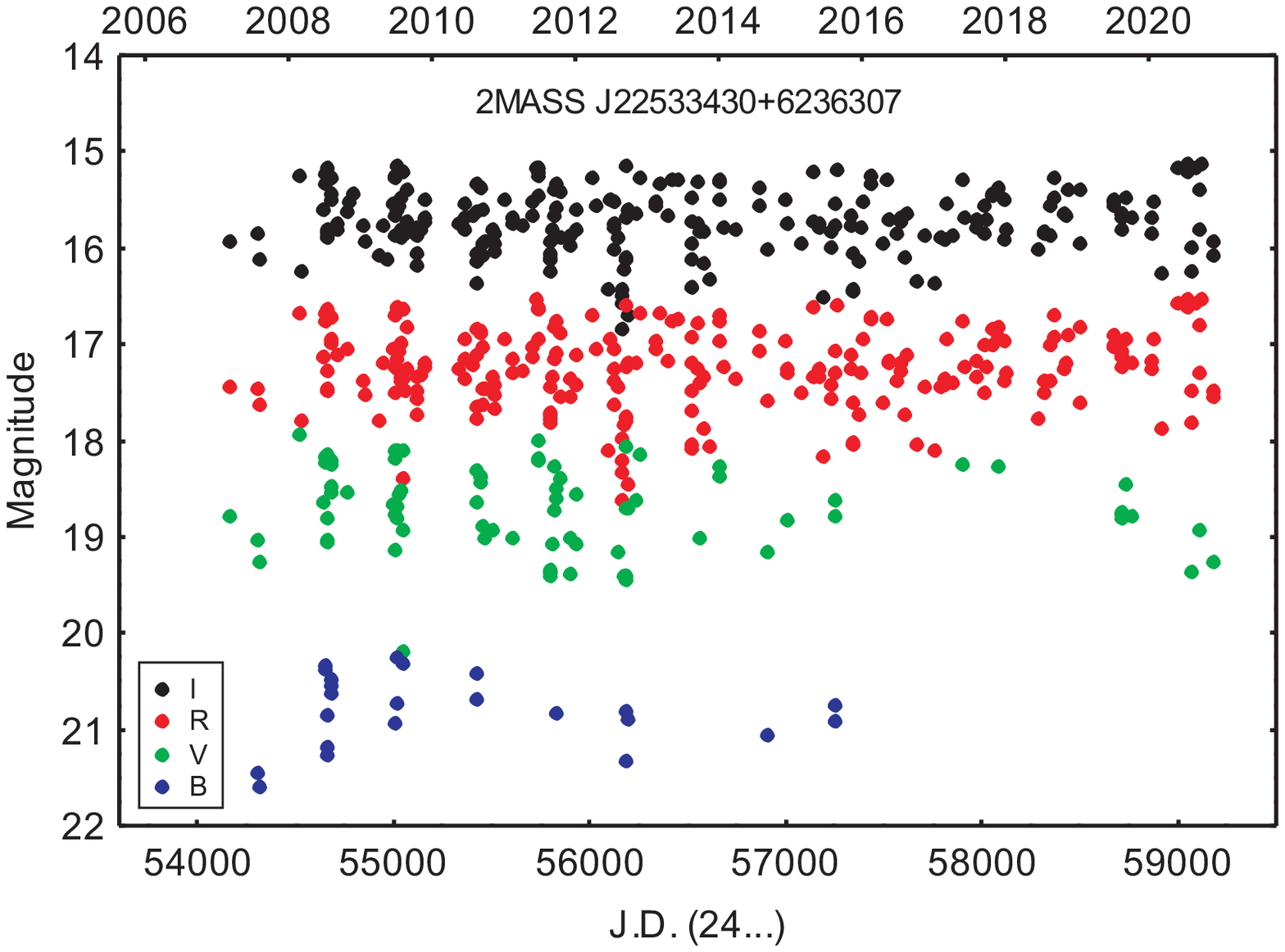}
		\includegraphics[width=7.55cm]{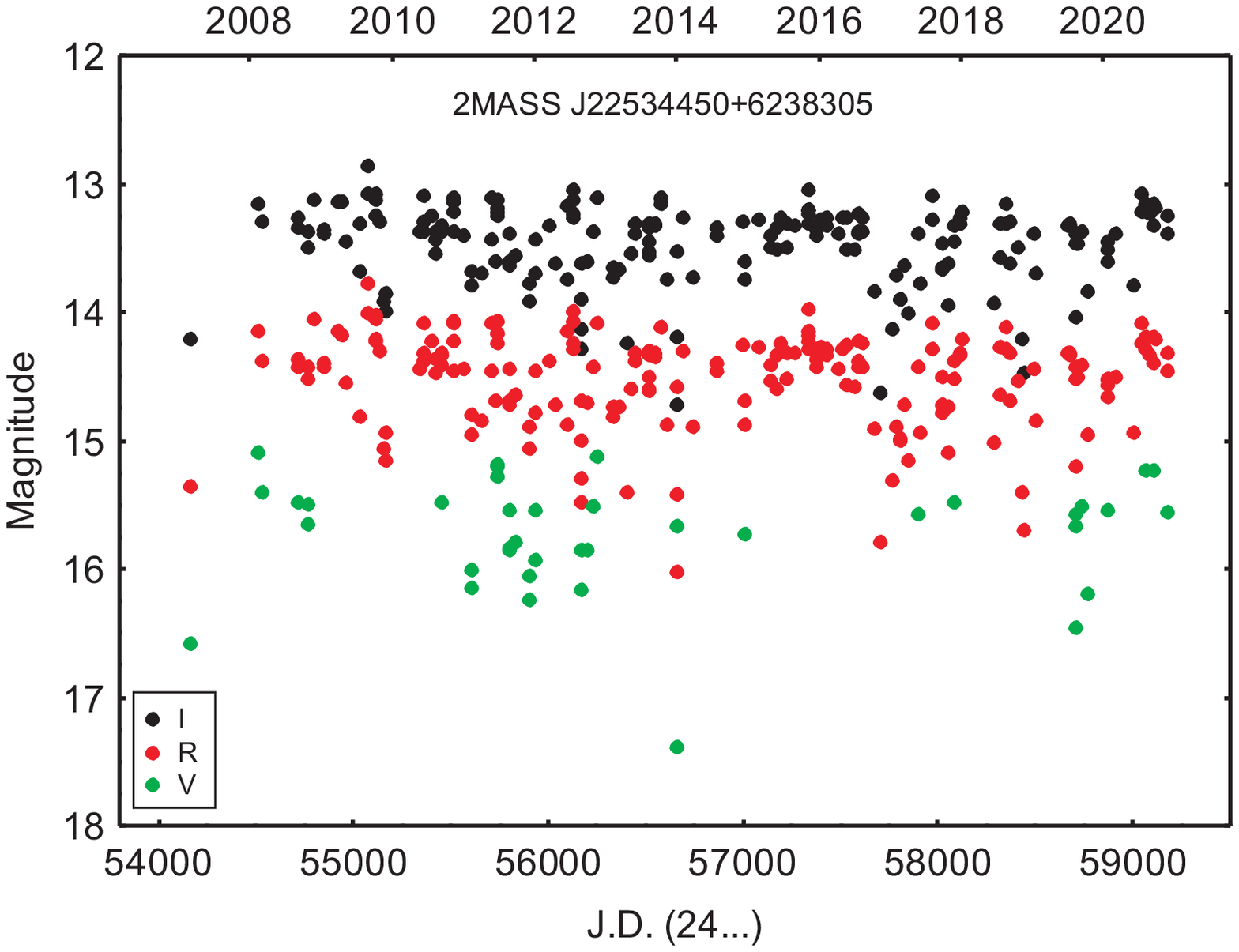}
		\includegraphics[width=7.55cm]{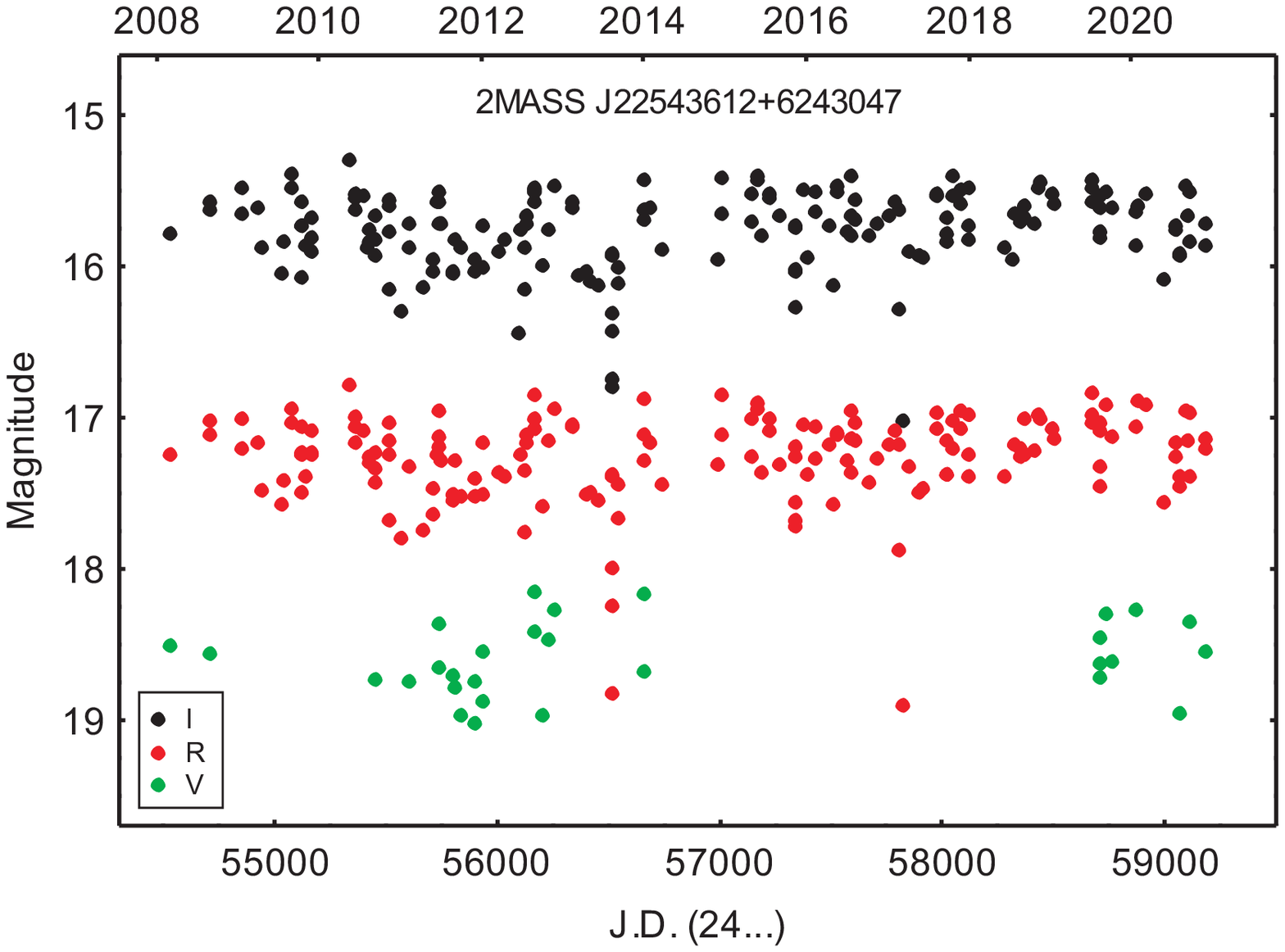}
		\includegraphics[width=7.55cm]{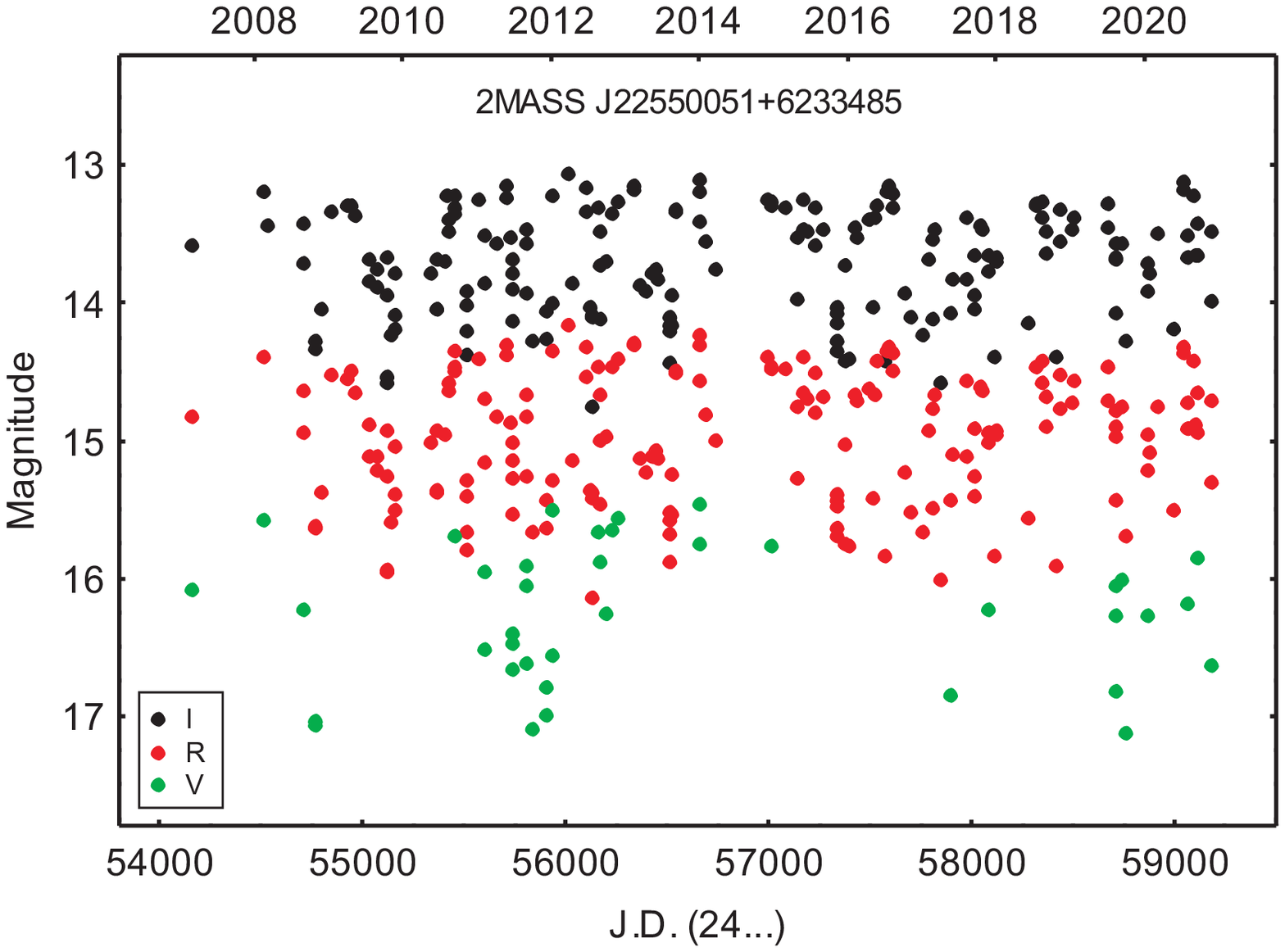}
		\includegraphics[width=7.55cm]{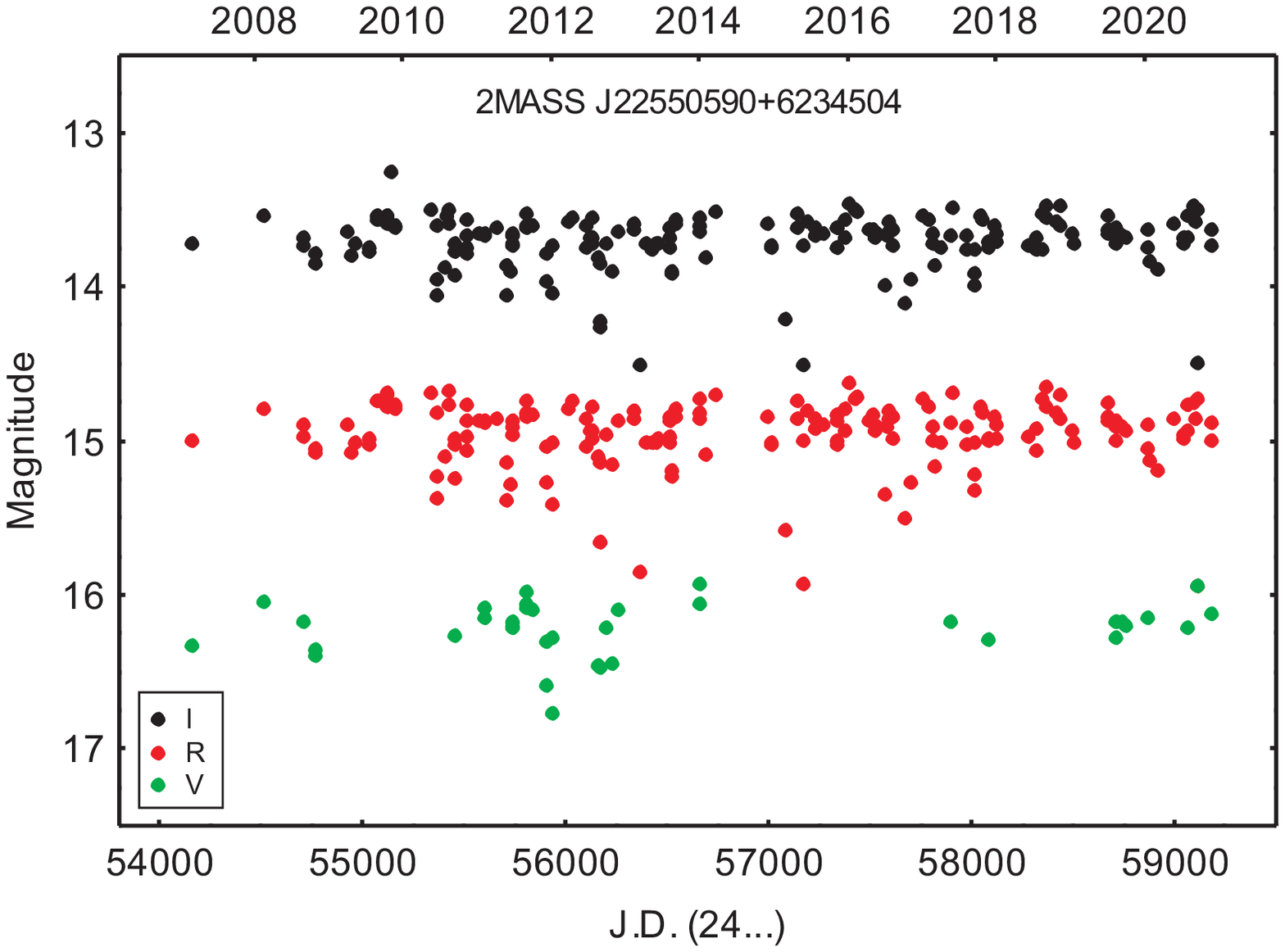}
		\includegraphics[width=7.55cm]{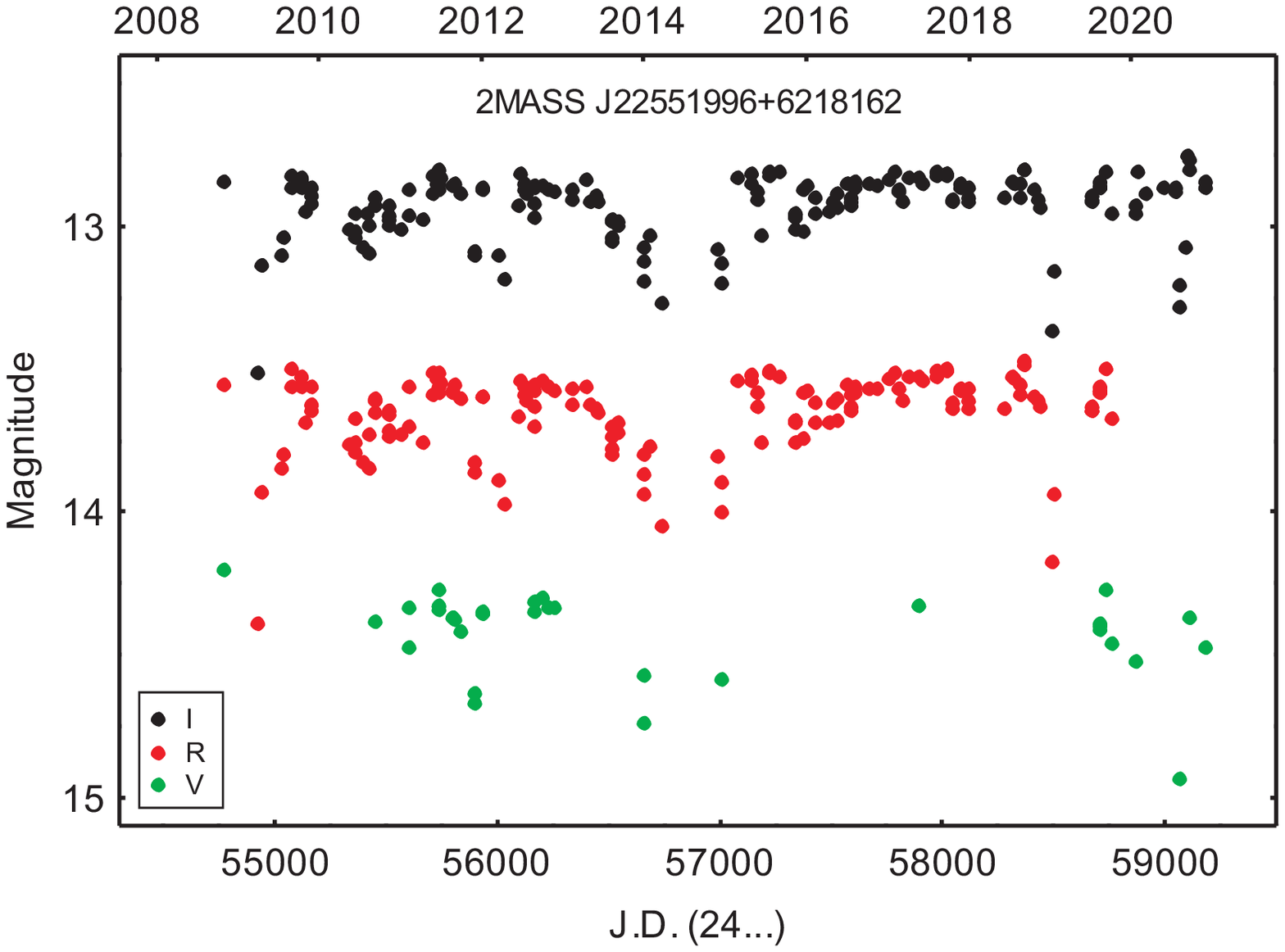}
		\includegraphics[width=7.55cm]{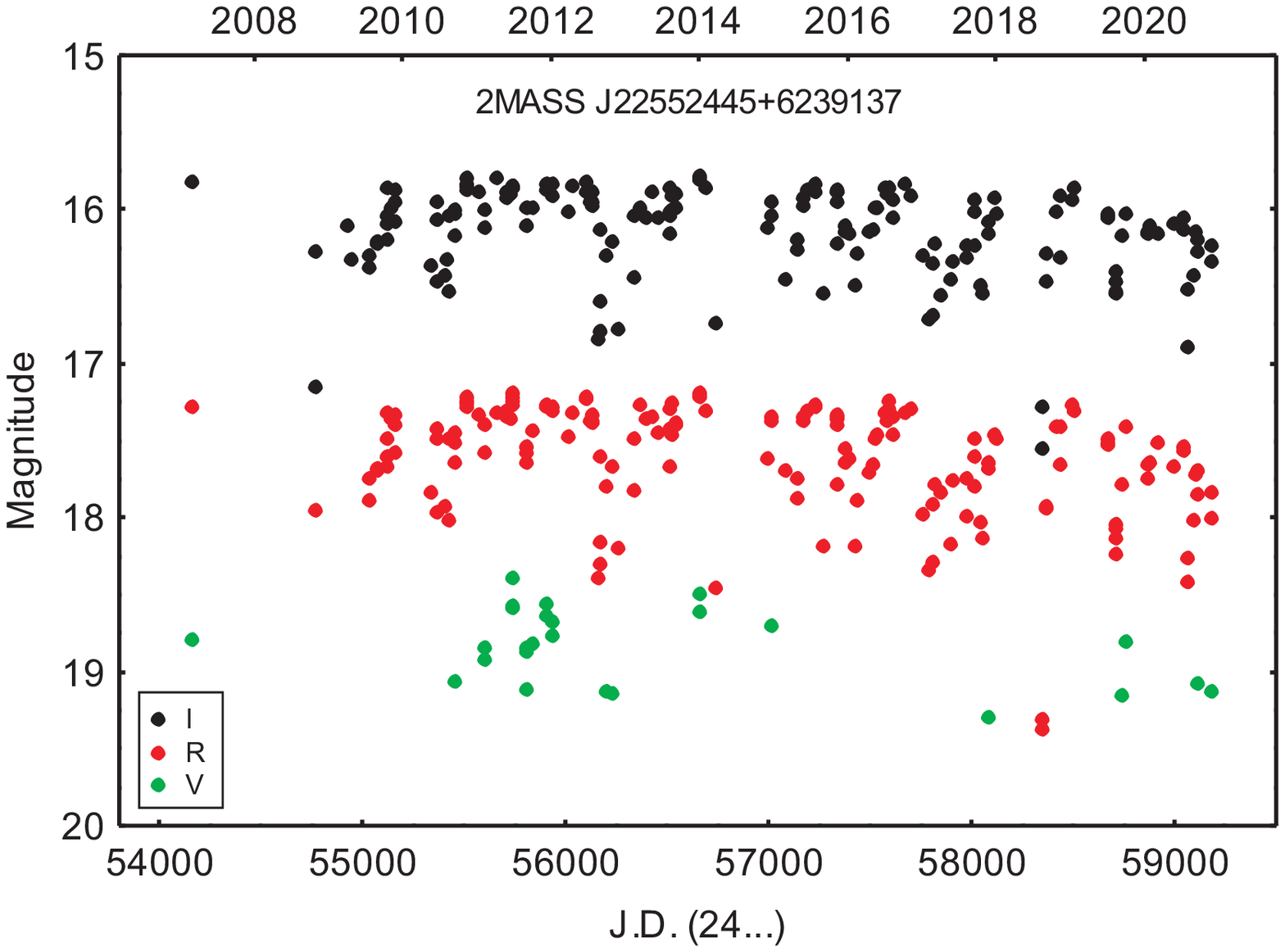}
		\caption{$V(RI)_{c}$ light curves of 2MASS J22533430+6236307, 2MASS J22534450+6238305, 2MASS J22543612+6243047, 2MASS J22550051+6233485, 2MASS J22550590+6234504, 2MASS J22551996+6218162 and 2MASS J22552445+6239137.}\label{Fig:curves3}
\end{figure}

The photometric behavior of these stars is characterized by strong variability and multiple brightness dips.
The light curves of the stars are shown in Fig.~\ref{Fig:curves3}.
The registered amplitudes of the brightness variations of the stars during our monitoring are given in Table~\ref{Tab:amplitudes}.
Usually, the large amplitude dips in brightness are an indication of UXor-type variability. 
In this case, it can be assumed that one of the reasons for the variability of these stars is the obscuration by circumstellar material.
On the other hand, the different shapes, amplitudes and durations of the observed dips give grounds to predict a variety obscuration reasons, e.g., clouds of proto-stellar material, massive dust clumps orbiting the central star, or planetesimals at different stages of formation.

It is seen from the color$-$magnitude diagrams (Fig.~\ref{Fig:colors}) that the stars become redder as they fade.
Such color variations are typical for PMS stars, whose variability is produced by irregular obscuration or the rotational modulation of one or more spots on the stellar surface.

\begin{figure}[h!]
	\centering
	\includegraphics[width=7cm, angle=0]{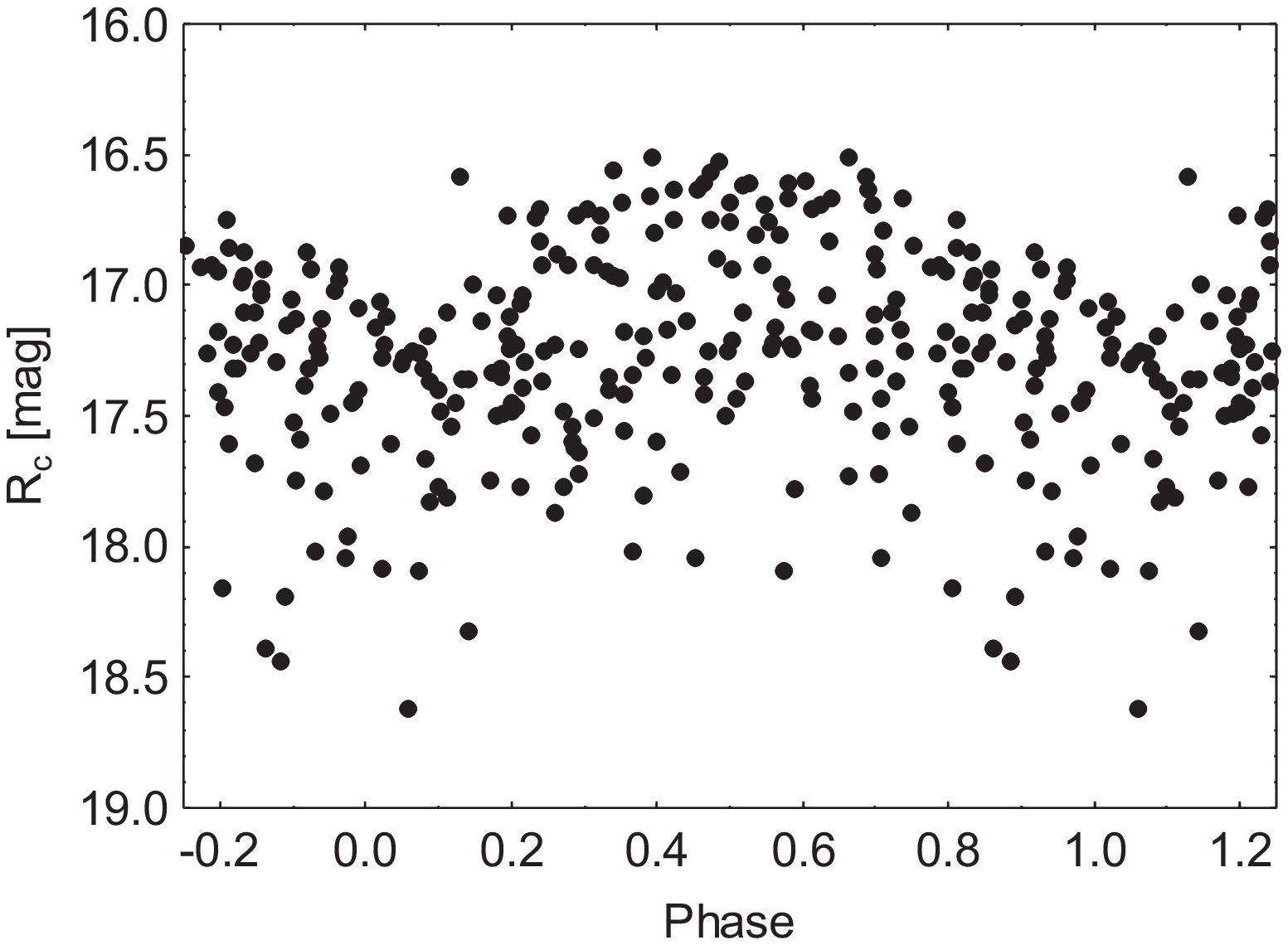}
	\caption{Phase-folded $R_{c}$ light curve of 2MASS J22533430+6236307 according to 11.68 day period.}\label{Fig:2MASSJ22533430+6236307_period}
\end{figure}

An important result from our study of 2MASS J22533430+6236307 is the identification of previously unknown periodicity in its brightness variation.
We found a significant peak in the star's periodogram corresponding to 11.68 day period.
This period remained stable during the whole time of our observations.
The phase-folded $R_{c}$ light curve of the star is displayed on Fig.~\ref{Fig:2MASSJ22533430+6236307_period}.

\subsection{2MASS J22541742+6236221}

\begin{figure}[h!]
	\begin{center}
		\includegraphics[width=8.5cm]{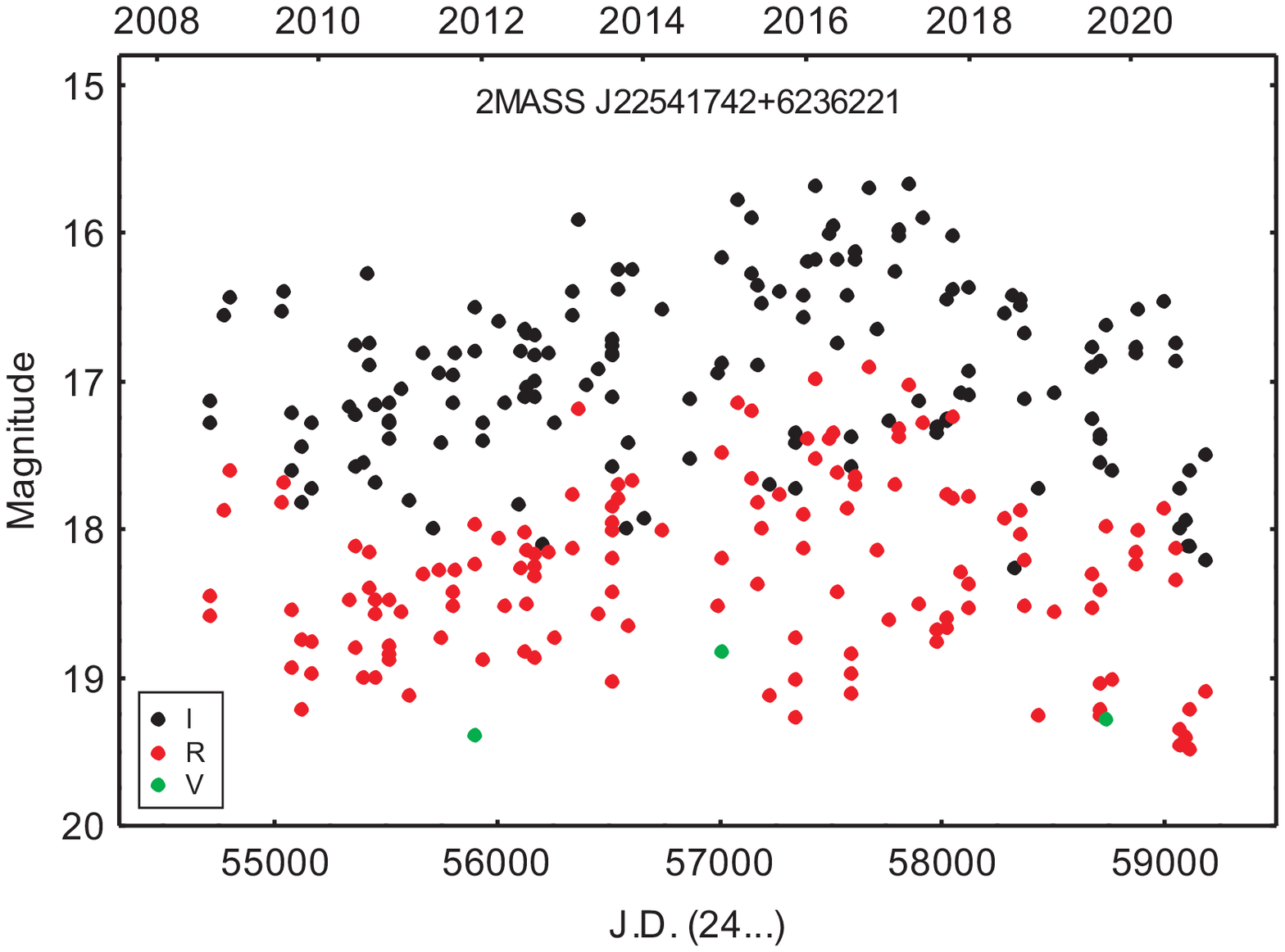}
		\caption{$V(RI)_{c}$ light curves of 2MASS J22541742+6236221.}\label{Fig:curves4}
	\end{center}
\end{figure}

During our observations, this star manifests photometric variability with the highest amplitude ($\Delta{I_{c}}$=2.60 mag).
In its light curves (Fig.~\ref{Fig:curves4}), both rapid increases and drops are observed.
Such strong variability could be caused by the highly variable mass accretion from the circumstellar disk onto the stellar surface.
Evidence of periodicity in the brightness variation of the star is not detected. 

\subsection{2MASS J22552178+6237535}

\begin{figure}[h!]
	\begin{center}
		\includegraphics[width=8.5cm]{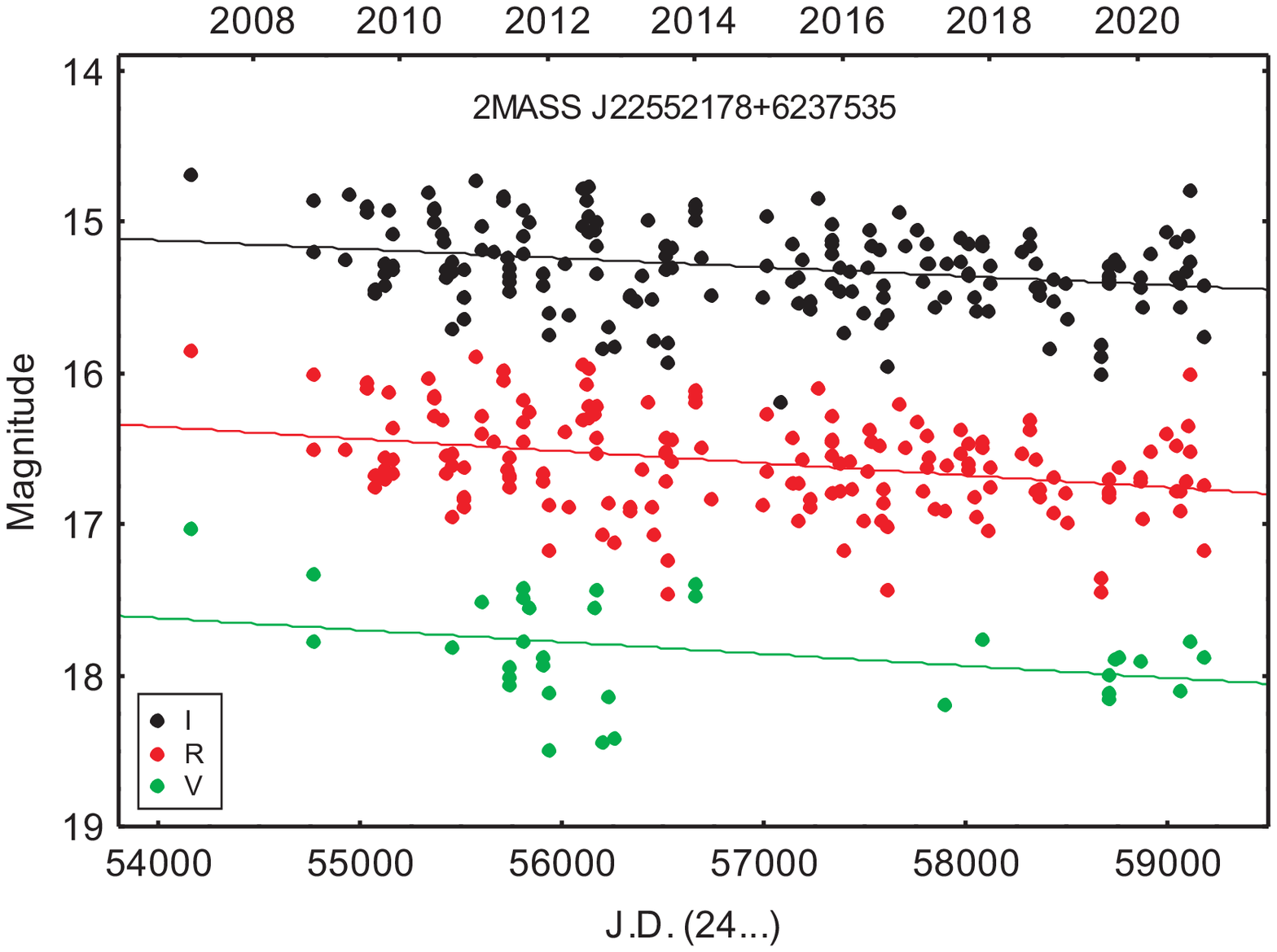}
		\caption{$V(RI)_{c}$ light curves of 2MASS J22552178+6237535.}\label{Fig:curves5}
	\end{center}
\end{figure}

This star exhibits irregular variability, whose amplitude is almost the same in different years.
Its light curves are depicted in Fig.~\ref{Fig:curves5}.
The registered amplitudes of the star's variability are given in Table~\ref{Tab:amplitudes}.

An important result from our study of 2MASS J22552178+6237535 is that the total star's brightness gradually decreases during the whole period of our observations.
Using a linear approximation for all data of the star, we calculated the following values for the rates of decreases: 22$\times$10$^{-2}$mag $yr^{-1}$ for the $I_{c}$ band, 29$\times$10$^{-2}$mag $yr^{-1}$ for the $R_{c}$ band, and 29$\times$10$^{-2}$mag $yr^{-1}$ for the $V$ band.
Evidence of reliable periodicity in the variability of 2MASS J22552178+6237535 is not detected.

\section{Conclusion}
\label{sect:conclusion remarks}

All monitored objects in the present study show typical for PMS stars strong photometric variability.
The obtained main results for the investigated stars can be summarized as follows:
(i) 2MASS J22515547+6218170 and 2MASS J22545815+6241327 exhibit brightness variations, whose amplitude is almost the same in different years;
(ii) 2MASS J22533129+6237114 and 2MASS J22551206+6228169 manifest large amplitude flare events;
(iii) the photometric behavior of 2MASS J22533430+6236307, 2MASS J22534450+6238305, 2MASS J22543612+6243047, 2MASS J22550051+6233485, 2MASS J22550590+6234504, 2MASS J22551996+6218162 and 2MASS J22552445+6239137 is characterized by strong variability and multiple brightness dips;
(iv) for 2MASS J22533430+6236307 is identified an 11.68 d periodicity in its light variation;
(v) the variability of 2MASS J22541742+6236221 likely is caused by the highly variable mass accretion;
(vi) the total brightness of 2MASS J22552178+6237535 decreases over time.

Our study adds 13 YSOs with long-term multicolor photometry to the PMS stars family.
Further regular observations would offer clearer insight into the physical nature of the investigated stars.
We plan to continue our photometric monitoring of the field of V733 Cep during the next years.

\begin{acknowledgements}
This research has made use of NASA's Astrophysics Data System Abstract Service.
This study has made use of the International Variable Star Index (VSX) database, operated at AAVSO, Cambridge, Massachusetts, USA.
The authors thank the Director of Skinakas Observatory Prof. I. Papamastorakis and Prof. I. Papadakis for the award of telescope time.
This research has made use of the VizieR catalogue access tool (Ochsenbein et al.~\cite{ochs00}), CDS, Strasbourg, France.
This work was partly supported by the National Science Fund of the Ministry of Education and Science of Bulgaria under grant DN 18-10/2017 and by funds of the project RD-08-125/2021 of the University of Shumen.
\end{acknowledgements}

\appendix                  


\begin{thebibliography}{28}
	
	\bibitem[2012]{alle12} Allen T. S., Gutermuth R. A., Kryukova E., et al., 2012, ApJ, 750, 125
	
	\bibitem[2018]{bail18} Bailer-Jones C. A. L., Rybizki J., Fouesneau M., et al., 2018, AJ, 156, 58

	\bibitem[1959]{blaa59} Blaauw A., Hiltner W. A., Johnson H. L., 1959, ApJ, 130, 69
	
	\bibitem[2020]{evit20} Evitts J. J., Froebrich D., Scholz A., et al., 2020, MNRAS, 493, 184
	
	\bibitem[2020]{froe20} Froebrich D., Scholz A., Eisl\"{o}ffel J., Stecklum B., 2020, MNRAS, 497, 4602	
	
	\bibitem[1991]{grin91} Grinin V. P., Kiselev N. N., Minikulov N. Kh., Chernova G. P., Voshchinnikov N. V., 1991, Ap\&SS, 186, 283
	
	\bibitem[2019]{hamb19} Hamb\'{a}lek L., Van\v{k}o M., Paunzen E., Smalley B., 2019, MNRAS, 483, 1642
	
	\bibitem[2007]{herb07} Herbst W., Eisl\"{o}ffel J., Mundt R., Scholz A., 2007, Protostars and Planets V, Eds. B. Reipurth, D. Jewitt \& K. Keil (University of Arizona Press, Tucson), 297
	
	\bibitem[2014]{huan14} Huang Y.-F., Li J.-Z., Rector T. A., Fan Z., 2014, RAA, 14, id. 1269-1278
	
	\bibitem[2015]{ibry15} Ibryamov S. I., Semkov E. H., Peneva S. P., 2015, \pasa, 32, id. e021
	
	\bibitem[2020]{ibry20} Ibryamov S., Semkov E., Peneva S., Gocheva K., 2020, RAA, 20, id. 194
	
	\bibitem[2005]{isma05} Ismailov N. Z., 2005, Astronomy Reports, 49, 309
	
	\bibitem[2005]{lenz05} Lenz P., Breger M., 2005, CoAst, 146, 53
	
	\bibitem[1999]{mena99} M\'{e}nard F., Bertout C., 1999, In: The Origin of Stars and Planetary Systems, Eds. C. J. Lada \& N. D. Kylafis (Kluwer Academic Publishers), 341
	
	\bibitem[2001]{mika01} Mikami T., Ogura K., 2001, Ap\&SS, 275, 441 
	
	\bibitem[1993]{more93} Moreno-Corral M. A., Chavarria-K. C., de Lara E., Wagner S., 1993, A\&A, 273, 619
	
	\bibitem[1999]{nayl99} Naylor T., Fabian A. C., 1999, MNRAS, 302, 714
	
	\bibitem[2000]{ochs00} Ochsenbein F., Bauer P., Marcout J., 2000, A\&AS, 143, 23
	
	\bibitem[2010]{pene10} Peneva S. P., Semkov E. H., Munari U., Birkle K., 2010, A\&A, 515, A24
	
	\bibitem[2003]{pozz03} Pozzo M., Naylor T., Jeffries R. D., Drew J. E., 2003, MNRAS, 341, 805
	
	\bibitem[2007]{reip07} Reipurth B., Aspin C., Beck T., et al., 2007, AJ, 133, 1000
	
	\bibitem[2017]{rigo17} Rigon L., Scholz A., Anderson D., West R., 2017, MNRAS, 465, 3889	
	
	\bibitem[1979]{sarg79} Sargent A. I., 1979, ApJ, 233, 163
	
	\bibitem[2019]{semk19} Semkov E. H., Ibryamov S. I., Peneva S. P., 2019, SerAJ, 199, 39
	
	\bibitem[2008]{semk08} Semkov E. H., Peneva S. P., 2008, IBVS, 5831, 1
	
	\bibitem[2019]{siwa19} Siwak M., Dr\'{o}\.{z}d\.{z} M., Gut K., Winiarski M., Og\l oza W., Stachowski G., 2019, AcA, 69, 227
	
	\bibitem[2006]{skru06} Skrutskie M. F., Cutri R. M., Stiening R., et al., 2006, AJ, 131, 1163
	
	\bibitem[1989]{vosh89} Voshchinnikov N V., 1989, Astrofizika, 30, 509
	
\end{thebibliography}
\end{document}